\documentclass[11pt]{article}
\pdfoutput=1
\usepackage{amsfonts}
\usepackage{amsmath,amssymb}
\usepackage{bbm}
\usepackage{bm}
\usepackage{color}
\usepackage{slashed}
\usepackage{cite}
\def\Ddouble{\overset{\text{\tiny$\bm\leftrightarrow$}}{\bm{\partial}}}
\def\DdoubleBIS{\overset{\text{\tiny$\bm\leftrightarrow$}}{\partial}}

\def\siml{{\ \lower-1.2pt\vbox{\hbox{\rlap{$<$}\lower6pt\vbox{\hbox{$\sim$}}}}\ }}
\def\simg{{\ \lower-1.2pt\vbox{\hbox{\rlap{$>$}\lower6pt\vbox{\hbox{$\sim$}}}}\ }}
\usepackage{stackengine}
\stackMath
\usepackage{graphicx}
\usepackage{bbold}
\DeclareGraphicsExtensions{.pdf,.png,.jpg}

\usepackage[colorlinks=false,citecolor=cyan,urlcolor=blue,bookmarks=true,bookmarks=true,bookmarksopen=true,bookmarksnumbered=true,bookmarksopenlevel=3]{hyperref}
\definecolor{airforceblue}{rgb}{0.36, 0.54, 0.66}
\definecolor{steelblue}{rgb}{0.27, 0.51, 0.71}
\definecolor{amber}{rgb}{1.0, 0.49, 0.0}
\def\lsi{\raise0.3ex\hbox{$<$\kern-0.75em\raise-1.1ex\hbox{$\sim$}}}
\def\gsi{\raise0.3ex\hbox{$>$\kern-0.75em\raise-1.1ex\hbox{$\sim$}}}

\newcommand{\gsim}{\mathop{\gsi}}

\allowdisplaybreaks[1]
\def\simg{{\ \lower-1.2pt\vbox{\hbox{\rlap{$>$}\lower6pt\vbox{\hbox{$\sim$}}}}\ }}
\def\siml{{\ \lower-1.2pt\vbox{\hbox{\rlap{$<$}\lower6pt\vbox{\hbox{$\sim$}}}}\ }}

\makeatletter \@addtoreset{equation}{section} \makeatother

\begin{document}

\flushbottom

\begin{titlepage}

\begin{centering}

\vfill

{\Large{\bf
   Coloured coannihilations:\\ Dark matter phenomenology meets non-relativistic EFTs
  }
}

\vspace{0.8cm}

S.~Biondini$^{\rm a}$ \footnote{s.biondini@rug.nl}
and S.~Vogl$^{\rm b}$ \footnote{stefan.vogl@mpi-hd.mpg.de}

\vspace{0.8cm}

{\em $^{\rm{a}}$Van Swinderen Institute, University of Groningen
 \\ 
Nijenborgh 4, NL-9747 AG Groningen, Netherlands} 
\\
\vspace{0.15 cm}
{\em $^{\rm{b}}$Max-Planck-Institut 
f\"ur Kernphysik (MPIK),\\ Saupfercheckweg 1, 69117 Heidelberg,
Germany}
\vspace*{0.8cm}

\end{centering}

\vspace*{0.3cm}
 
\noindent
\textbf{Abstract}: We investigate the phenomenology of a simplified model with a Majorana fermion as dark matter candidate which interacts with Standard Model quarks via a colour-charged coannihilation partner. Recently it has been realized that non-perturbative dynamics, including the Sommerfeld effect, bound state formation/dissociation and thermal corrections, play an important role in coannihilations with coloured mediators. This calls for a careful analysis of thermal freeze-out and a new look at the experimental signatures expected for a thermal relic. We employ a state of the art calculation of the relic density which makes  use of a non-relativistic effective
theory framework and calculate the effective annihilation rates by  solving a plasma-modified Schr\"odinger equation. We determine the cosmologically preferred parameter space  and  confront it with current experimental limits and future prospects for dark matter detection.  

\vfill

\vfill
\newpage
\end{titlepage}

\section{Introduction}
Astrophysical and cosmological observations clearly show that the Standard Model of particle physics (SM) describes only about $20\%$ of the matter content of the Universe. The remaining $\approx80\%$ can be described to excellent precision by a new cold and collisionless kind of matter.  Understanding the nature and the origin of this dark matter (DM) is one of the most pressing issues in high energy physics and cosmology.

Weakly interacting massive particles (WIMPs) produced by thermal freeze-out in the early Universe are arguably one of the most appealing candidates for DM. In this picture the DM has been in thermal equilibrium with the SM plasma at high temperatures. While the Universe cools down, the annihilations rate of DM pairs into SM particles cannot keep up with the expansion rate, and DM  finally decouples from the thermal environment. However, the great improvements of direct detection experiments, which exclude cross sections as low as 
$4.1 \times 10^{-47}\, \mbox{cm}^2$~\cite{Aprile:2018dbl},
lead to an increasing tension between experimental results and theoretical expectations in models in which the direct detection and annihilation cross sections are related by a crossing symmetry~\cite{Duerr:2016tmh,Escudero:2016gzx,Athron:2017kgt}.   

There exist a number of well-known scenarios in which this crossing symmetry is not realized. Consequently, the correlation between the annihilation rate and the direct detection rate is more complicated than naive estimates might suggest. One particularly interesting possibility is that the DM is not the only state in the dark sector which is present during freeze-out. In this case ``coannihilation'', i.e. processes which include additional degrees of freedom from the dark sector in the initial or final state, play a role in setting the relic density \cite{Griest:1990kh,Edsjo:1997bg}. The thermally averaged (co)annihilation rates which control the evolution of the DM in the early Universe get suppressed very rapidly once the mass difference between the DM and the coannihilation partner is comparable to the temperature $T$ at freeze-out. Therefore, only coannihilation partners with a comparatively large interaction rate or a very strong mass degeneracy with the DM allow for efficient coannihilations.
This makes colour charged mediators, which possess a large QCD annihilation cross section, popular coannihilation partners for the DM \cite{deSimone:2014pda,Ellis:2014ipa,Ibarra:2015nca,Pierce:2017suq,Baker:2015qna,ElHedri:2017nny,Davoli:2018mau,Garny:2018icg}.

The phenomenology of coloured mediators is extremely rich. On the one hand, it is well established that non-perturbative effects ranging from Sommerfeld-enhancement to bound-state formation play a non-negligible role in DM production for coloured coannihilations \cite{deSimone:2014pda,vonHarling:2014kha,Ibarra:2015nca,Harz:2014gaa,Liew:2016hqo,Kim:2016kxt,Kim:2016zyy}. As freeze-out occurs in the thermal bath of the hot and early Universe, a careful consideration of the total annihilation rate which takes non-perturbative effects in a thermal background into account is crucial for a reliable determination of the relic density. In particular, the inclusion of bound-state
effects for DM annihilation has recently been shown to have a large impact on the relic abundance and increases the largest DM mass compatible with the observed DM density considerably \cite{Liew:2016hqo,Mitridate:2017izz,Biondini:2018pwp,Harz:2018csl}. On the other hand, coloured mediators can be tested  at the LHC, where their QCD interactions allow for copious production,  and direct detection experiments. In this paper we try to connect theses different  aspects and combine state-of-the-art predictions for thermal freeze-out with a detailed study of the phenomenology at colliders and direct detection experiments. In this way we can confront the cosmological expectations for a thermal relic with other observables and map out the parts of the theoretical favoured parameter space which are in agreement with laboratory searches.

This article is organized as follows. First,  we introduce a simplified model for coannihilation which we use for our calculations. In Sec.~\ref{Sec:RelicDensity} we discuss the computation of the relic density with a particular focus on the challenges which arise due the presence of a colour charged coannihilation partner in the plasma in the early Universe.
Next, we discuss experimental observables which allow to test the model under consideration with present and future detectors in Sec.~\ref{sec:exp_constraint}. We map the results of laboratory experiments on the theoretically favoured parameter space of a thermal relic in Sec.~\ref{Sec:Results}. Finally, conclusions and outlook are offered in Sec.~\ref{Sec:Conclusions}. Some technical details regarding the non-relativistic operators accounting for velocity suppressed annihilation cross section and the thermal potentials used in the analysis are given in the  Appendix \ref{Appx:higherOperators} and \ref{Appx:thermalpotential} respectively. 
\section{Simplified Model}
The simplified model that we consider consists of
a gauge singlet Majorana fermion ($\chi$) and a scalar field
($\eta$), the latter 
is a singlet under SU$_L$(2) but carries non-trivial QCD 
and hypercharge quantum numbers. In the MSSM framework, 
the Majorana fermion can be identified with a bino-like neutralino and the scalar 
with a right-handed stop or more generally any right-handed squark. However,
we do not fix couplings to their MSSM values and treat them as free parameters.

The Lagrangian for this extension of the Standard Model can be 
expressed as \cite{Garny:2015wea} 
\begin{eqnarray}
 \mathcal{L} & = & 
 \mathcal{L}^{ }_{\hbox{\tiny SM}} + 
 \frac{1}{2} \, \bar{\chi} \left(  i \slashed{\partial} - M_\chi \right)  \chi 
 + (D^{ }_\mu \eta)^\dagger D^\mu \eta 
 - M_\eta^2\, \eta^\dagger \eta 
 - \lambda^{ }_2 (\eta^\dagger \eta)^2 
 \nonumber 
 \\
 & - & \lambda^{ }_3\, \eta^\dagger \eta\, H^\dagger H 
 - y\,  \eta^\dagger \bar{\chi} P_R q 
 - y^* \bar{q} P_L \chi\, \eta
 \;,
 \label{Lag_RT}
\end{eqnarray}
where $H$ is the Standard Model Higgs doublet, $M_\eta$ the mass of the mediator and $M_\chi$ the mass of the DM particle. The Yukawa coupling between $\eta$ and $\chi$ is denoted by $y$ while $\lambda_2$ and $\lambda_3$  are the self-coupling of the coloured scalar and its coupling to the Higgs, respectively.  The coupling $\lambda_1$ is left for the Standard Model Higgs self interaction and $P_R$ ($P_L$) is the right-handed (left-handed) projector. 
\section{Deriving the relic density}
\label{Sec:RelicDensity}
With the precise observation of the Cosmic Microwave Background (CMB) by the Planck satellite the cosmological abundance of DM has been measured to be $\Omega_{\hbox{\tiny DM}} h^2 = 0.1200(12)$~\cite{Akrami:2018vks}. Given that this is by far the most precise measurement of DM, an accurate theoretical prediction of the relic density is key ingredient of any realistic study of DM phenomenology. In this context it is crucial to determine the degrees of freedom which are relevant for the production of the DM in the early Universe. In models with coloured mediators 
the relative abundance of $\eta$ and $\chi$ is typically proportional to the ratio of their equilibrium  densities since the conversion rates between them are  much larger than the Hubble rate over a large range of temperature and chemical equilibrium holds during the freeze-out process.\footnote{ See however \cite{Garny:2017rxs} for a discussion of the dynamics in the absence of chemical equilibrium.}
Consequently, two regimes can be distinguished: (i) for large values of $\Delta M = M_\eta -M_\chi$ the abundance of coloured mediators in the plasma is highly suppressed during the freeze-out of DM and established methods for the computation of the relic density are readily applicable (ii) for $\Delta M/ M \lesssim 0.2$ the abundance of $\eta$ is non-negligible and can influence the dynamic during freeze-out.  

The freeze-out mechanism sets the abundance of the Majorana DM fermions and coloured scalars. The picture is the standard one for heavy particles annihilating in pairs in a thermal ensemble: starting with an equilibrium distribution, the chemical equilibrium is gradually lost when the temperature drops below the heavy particle mass $T \ll M_\chi$ and recombination processes become Boltzmann suppressed. Eventually the production and annihilation rates cannot keep up with the expansion rate of the Universe and the heavy particles become so rare that they decouple from the thermal medium. The coloured scalars are non-relativistic objects around the freeze-out and during later stages of the annihilations. Indeed kinetic equilibration makes their velocity $v \sim \sqrt{T/M}$ to be smaller than unity. In the case of coannihilating species close in mass with the actual DM particle, the system of Boltzmann equations can be simplified and the evolution of the whole system can be traced with a single effective Boltzmann equation   \cite{Lee:1977ua,Gondolo:1990dk,Griest:1990kh}
\begin{equation}
\frac{dn}{dt} + 3 Hn =-\langle \sigma_{{\rm{eff}}} v \rangle (n^2-n^2_{{\rm{eq}}}) \, .
\label{BE_gen}
\end{equation}
The total equilibrium number density, which accounts for both particle species of the dark sector ($\chi$ and $\eta$), is
\begin{equation}
n_{{\rm{eq}}}= \int_{\bm{p}} e^{-E_{\bm{p}}/T} \left[ 2 + 2N_c \, e^{-\Delta M_T/T} \right] \, , 
\end{equation}
 and the effective annihilation cross section reads 
\begin{equation}
\langle \sigma_{{\rm{eff}}} v \rangle = \sum_{i,j} \frac{n^{\hbox{\scriptsize eq}}_i \,  n^{\hbox{\scriptsize eq}}_j}{(\sum_k n_k^{\hbox{\scriptsize eq}})^2} \langle \sigma_{ij} v  \rangle \,.
\label{co_cross}
\end{equation}
The mass splitting $\Delta M_T$ comprise both the in-vacuum and thermal contributions \cite{Biondini:2018pwp}, the latter originated from the interactions with particles in the plasma and $N_c=3$. The sum in eq.~(\ref{co_cross}) extends over all coannihilating particles, namely $i=\chi, \eta$. The key ingredient is the thermally averaged annihilation cross section in eq.~(\ref{BE_gen}) which controls the  dynamics in the early Universe. Normally, $\langle \sigma_{{\rm{eff}}} v \rangle$ is determined by averaging the in-vacuum cross sections $\sigma_{ij} v$ over the center-of-mass energies in a thermal plasma and reweighting the results according to eq.~(\ref{co_cross}).
However, the situation can be more complicated and subtle if the coannihilating species can interact with particles from the thermal bath. In our case, the scalar particles feel QCD strong interactions with quark and gluons and the thermally average cross section exhibits a rather strong temperature dependence. Therefore, the in-vacuum result is not a good approximation~\cite{Liew:2016hqo,Kim:2016kxt,Kim:2016zyy,Mitridate:2017izz,Keung:2017kot,Biondini:2018pwp,Harz:2018csl}. In particular bound-state formation is especially efficient at small temperatures and make late-stage annihilations important.   
\begin{figure}[t!]
\centering
\includegraphics[scale=0.57]{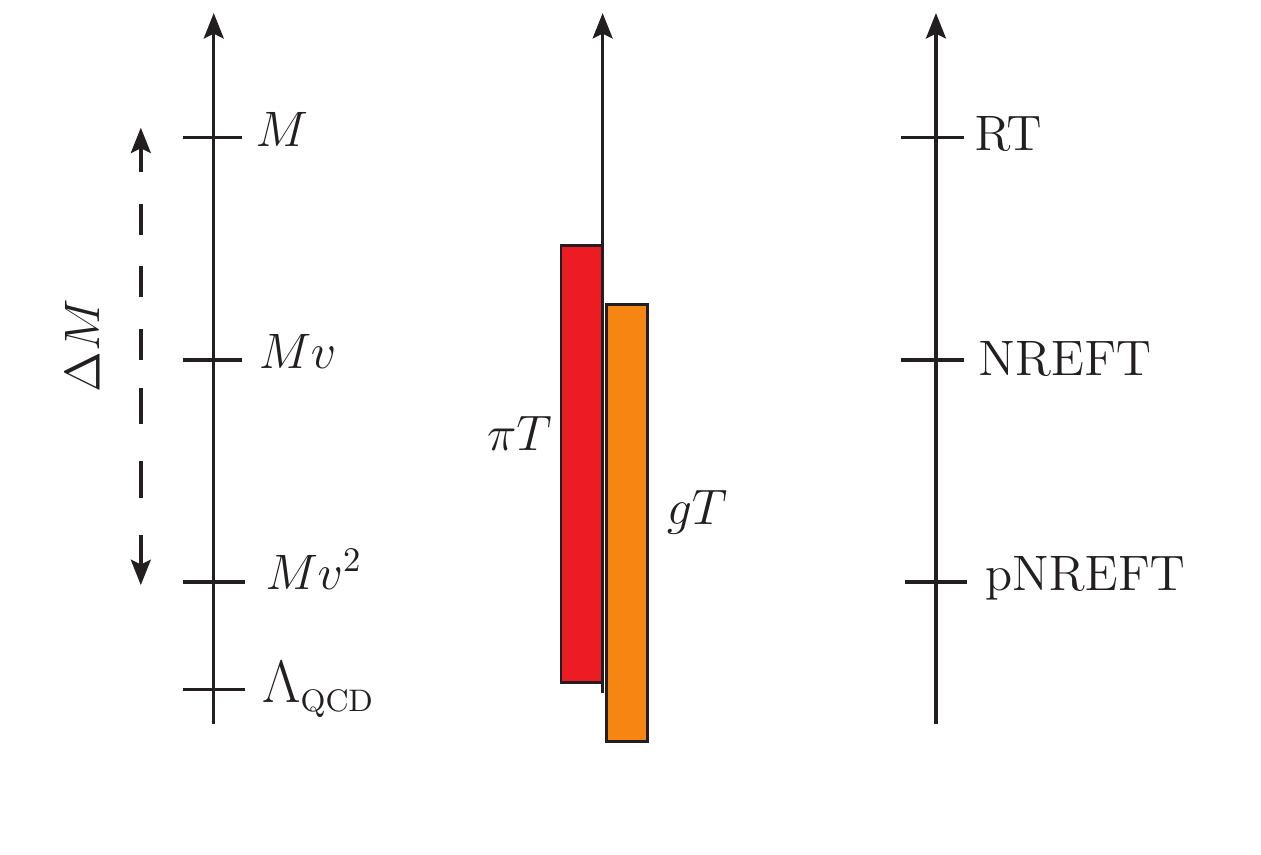}
\hspace{2.5 cm}
\includegraphics[scale=0.52]{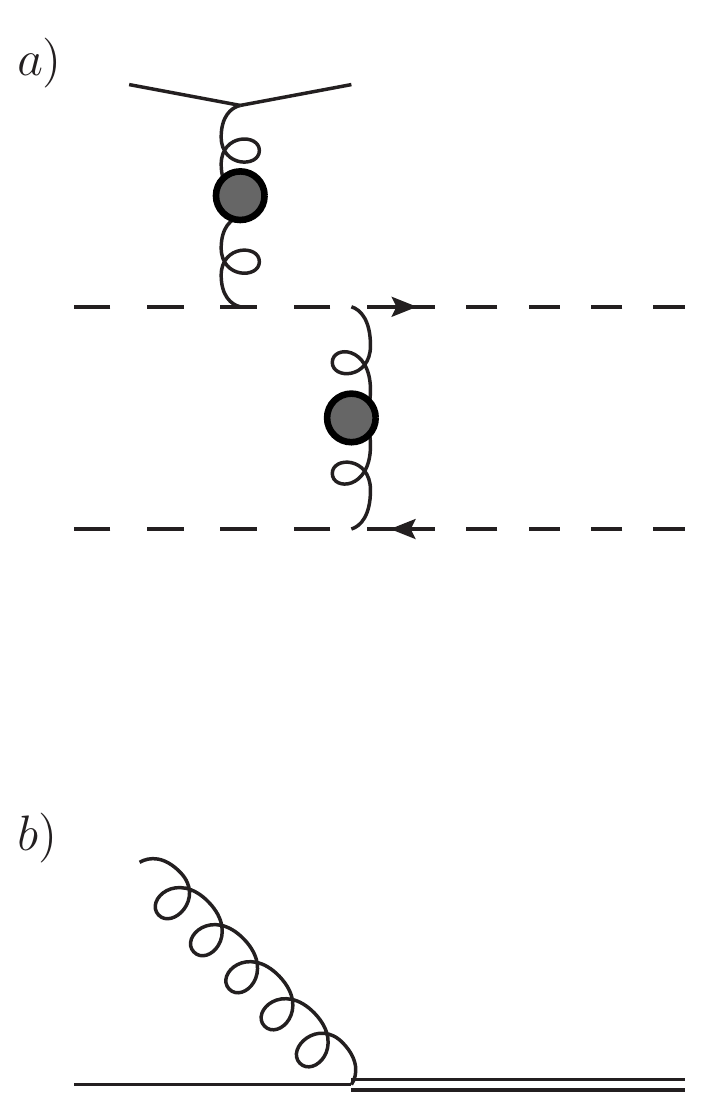}
\caption{\label{fig:figEFT}Left: non-relativistic and thermal scales. In the weak coupling regime the thermal scales are arranged as $\pi T \gg m_D$. Figure adapted from ref.\cite{Vairo:2015uva}, RT stands for relativistic theory. Right: Landau damping and gluodissociation processes, i.e. $2 \to 2$ scattering process with light plasma constituents and $2 \to 1$ process that involves a thermal gluon. In diagram a), dashed lines stand for coloured scalars and the gray blob for a resummed HTL gluon propagator. In diagram b), the solid line stands for a coloured scalar pair in the singlet configuration whereas the double solid line for a colour octet configuration. }
\end{figure}

Upon the standard definition of the yield parameter $Y \equiv n/s$, where $s$ is the entropy density, and  the change of variable
from time to $z \equiv M_\chi/T$, one can rewrite the Boltzmann equation and obtains
\begin{equation}
Y'(z)= - \langle  \sigma_{{\rm{eff}}} v \rangle M_\chi m_{{\rm{Pl}}} \times \frac{c(T)}{\sqrt{24 \pi e(T)}} \times \left.  \frac{Y^2(z)-Y_{\hbox{\scriptsize eq}}^2(z)}{z^2} \right|_{T=M_\chi/z} \, ,
\label{Boltzmann_Y}
\end{equation}
where $m_{{\rm{Pl}}}$ is the Planck mass, $e$ is the energy density, and $c$ is the heat capacity, for which we
use values from ref. \cite{Laine:2015kra}. We scan over the model parameters $y \in [0.1,2]$ and $\lambda_3 \in [0,1.5]$ and for every combination $(y, \lambda_3)$ we impose the condition $\Omega_{\hbox{\tiny DM}} h^2 =0.1200(12)$\cite{Akrami:2018vks}. This way a relation between the DM mass and mass splitting is obtained to constrain the model. 
\subsection{Coloured-scalar dynamics in a thermal bath}
The coloured scalars are moving slowly  and they can undergo several interactions before annihilating into light Standard Model particles. For example, multiple exchange of gluons leads to the Sommerfeld effect that increase (decrease) the annihilation rate for an attractive (repulsive) potential experienced by the heavy pair. Moreover, the coloured scalar can potentially form bound states in the medium. As the two particles are close to each other in the bound state, their annihilations become more efficient and reduce the relic density. The Majorana fermion does not interact with gluons at tree level, therefore long-range interaction do not appear in $\chi \chi$ annihilations or $\chi \eta$ coannihilation processes.  

Treating the dynamics of colour charged particles  in a medium (including bound-state dynamics formation and dissociation) is not a trivial task.
Even though we are interested in coloured scalars in this work, the problem is quite similar to heavy quarkonium in medium which can be described as two quasi-static colour sources sitting in a thermal ensemble. Profiting from the recent progress in quarkonium physics at finite temperature \cite{Laine:2006ns, Brambilla:2008cx,Brambilla:2010vq} we adopt an effective field theory (EFT) approach which is inspired by these results.
 In this framework, tailored quantum field theories are derived to effectively study the physics at a given energy/momentum scale. Two sets of energy scales play a role. On the one hand,  the non-relativistic scale $M_\eta$, the typical momentum $M_\eta v$ and kinetic energy $M_\eta v^2$ with the  hierarchy $M_\eta \gg M_\eta v \gg M_\eta v^2$ are relevant. On the other hand, there are thermal scales such as the plasma temperature $\pi T$ and the Debye mass $m_D$ which accounts for the chromoelectric-screening inverse length, or equivalently, a thermal mass acquired by the gluons.\footnote{$\pi T$ stands for the temperature scale where the factor $\pi$ is a remnant of the Matsubara modes of thermal field theory. For coulombic and near-coulombic states, the velocity is $v \sim \alpha_s$. So the momentum and binding energy/kinetic energy can be written as $M_\eta \alpha_s$ ad $M_\eta \alpha_s^2$ for the scale estimation that it more appropriate for a bound state, whereas those in terms of velocities better describe scattering states.}   A schematic arrangement of the energy scales is shown in figure~\ref{fig:figEFT}.  
  In our case, we are after the dynamics at the binding energy (and kinetic energy) scale, identified with $M_\eta v^2$, and its interplay with the thermal scales.
  
  Based on the finding for heavy quarkonium the following in-medium effects are expected to be relevant for coloured scalars: the Coulomb potential is modified by thermal effects and, more importantly, an imaginary part arises in the potential that corresponds to two different processes, namely the Landau damping and  thermal break up from bound to unbound pair triggered by a gluon of the plasma~\cite{Laine:2006ns, Brambilla:2008cx}. 
  The first process involves $2 \to 2$ scatterings with light particles from the medium whereas the second one describes the absorption of a thermal gluon. In a non-EFT framework, they are known as gluo-dissociation \cite{Kharzeev:1994pz} and dissociation by inelastic parton scattering \cite{Grandchamp:2001pf}. The equivalence of the  thermal breakup process to gluo-dissociation
has been analysed in \cite{Brambilla:2011sg}, whereas the relation between
the Landau-damping mechanism and the dissociation by inelastic parton scattering has been investigated in ref.\cite{Brambilla:2013dpa}. We sketch the two processes in figure~\ref{fig:figEFT}.  
So far, only one process at a time has been included in the bound-state dynamics and relic density determination even though both processes are known to be active in a thermal environment. In the following analysis, we shall consider both processes in the attempt to attain a more comprehensive description of bound-state effects for this model.

In this work, we adopt a formalism suited for addressing
the thermal annihilation of non-relativistic particles \cite{Kim:2016zyy,Kim:2016kxt}. The core of the method is to interpret $\langle \sigma_{{\rm{eff}}} v \rangle$ as a chemical equilibration rate $\Gamma_{{\rm{chem}}}$, which can be defined either on the perturbative or non-perturbative
level within linear response theory \cite{Bodeker:2012gs}. In this language $\Gamma_{{\rm{chem}}}$ can be traced back to the thermal expectation values of four-particle operators that describe heavy particle annihilations in a non-relativistic EFT (NREFT). In the following, we review the approach with a focus on the model under consideration and the connection with NREFTs and potential NREFTs (pNREFTs).

\subsection{Generalized Sommerfeld factors from NREFT and spectral functions}
\label{sec:NREFT_spectr}

The early Universe sets the stage for heavy-particle annihilations.  The freeze-out dynamics starts off when the temperature of the thermal bath drops the DM mass. The Majorana fermions and coloured scalars can be considered heavy because their mass is larger than any other energy scale in the problem. In particular, their typical momentum $M_{\chi(\eta)} v$ and kinetic energy $M_{\chi(\eta)} v^2$ are smaller than the mass. The same stands for the plasma temperature in the regime of interest. In other words, momentum modes as large as the DM mass are highly virtual and irrelevant for the non-relativistic dynamics in a medium with  $M_\chi \gg T$. Hence, the degrees of freedom are thermalized Standard Model particles with momentum of order $T$ and non-relativistic $\eta$ and $\chi$ particles (they both have similar masses in our setting). In order to obtain the corresponding EFT, one has to integrate out hard energy/momentum modes of order $M_\chi$ by setting all the other scales to zero, including the thermal scales.\footnote{The mass splitting is a small energy scale that can be still dynamical in the non-relativistic EFTs. Then, strictly speaking, we said that $M_\chi$ is integrated out, and both the species in the dark sector are non-relativistic because they are assumed to be close in mass.} As a result, the matching between the fundamental theory (\ref{Lag_RT}) and the non-relativistic theory can be done as in-vacuum. The so-obtained low-energy theory describes non-relativistic Majorana fermions and coloured scalars interacting with light degrees of freedom. As far as the coloured particles are concerned, the prototype for such EFT is non-relativistic QCD (NRQCD) describing heavy quarks \cite{Bodwin:1994jh}. An EFT describing non-relativistic Majorana fermions has been formulated in refs.\cite{Kopp:2011gg,Biondini:2013xua}.  As typical of non-relativistic field theories, the number of heavy particles is exactly conserved. Since light degrees of freedom with momenta of the order of the large mass $M_\chi$ have been integrated out, it seems we cannot describe accurately heavy-particle annihilations into final state particles with hard momenta. However, the inclusive annihilation rate can be recast  in terms of an
amplitude that conserves the number of the heavy particles thanks to the optical
theorem\cite{Bodwin:1994jh,Braaten:1996ix}. Therefore, particle-antiparticle annihilations are accounted for by introducing four-particle operators.  The operators are organized in a $1/M^2_\chi$ expansion, where the small parameter is the velocity of the heavy particles. Such an expansion resembles the velocity expansion of the annihilating states ($s$-wave, $p$-wave and so on). The matching coefficients of these operators contain the effects of the relativistic degrees of freedom and are obtained by matching four-point Green's functions between the fundamental and the low-energy theory, here the NREFT.

Starting from the relativistic theory written in (\ref{Lag_RT}) and expanding in the non-relativistic limit the scalar and fermion fields, the Lagrangian that comprise the four-particle dimension-6 operators read\cite{Biondini:2018pwp}
\begin{eqnarray}
 \mathcal{L}_{\hbox{\tiny NREFT}}^{d=6} & = & 
 i \, \Bigl\{ 
 \frac{c^{ }_1}{M_\chi^2} \, 
 \psi^\dagger_p \psi^\dagger_q \psi^{ }_q \psi^{ }_p 
 + 
 \frac{c^{ }_2}{M_\chi^2} \, 
 \bigl(
   \psi^\dagger_p \phi^\dagger_\alpha \psi^{ }_p \phi^{ }_\alpha +  
   \psi^\dagger_p \varphi^\dagger_\alpha \psi^{ }_p \varphi^{ }_\alpha   
 \bigr)  + 
 \frac{c^{ }_3}{M_\chi^2} \,  
\phi^\dagger_\alpha \varphi^\dagger_\alpha \varphi^{ }_\beta \phi^{ }_\beta
 \nonumber 
 \\
\phantom{xxx}& + &
 \frac{c^{ }_4}{M_\chi^2} \,  
\phi^\dagger_\alpha  \varphi^\dagger_\beta\,
 \varphi^{ }_\gamma \phi^{ }_\delta
 \, T^{a}_{\alpha\beta} T^{a}_{\gamma\delta}
  + 
 \frac{c^{ }_5}{M_\chi^2} \, 
 \bigl( \phi^\dagger_\alpha \phi^\dagger_\beta
        \phi^{ }_\beta \phi^{ }_\alpha 
   +    \varphi^\dagger_\alpha \varphi^\dagger_\beta
        \varphi^{ }_\beta \varphi^{ }_\alpha 
 \bigr)
 \Bigr\} \, .
 \label{Lag_NREFT}
\end{eqnarray}
where $\phi$, $\varphi$ and $\psi$ annihilate a scalar particle, a scalar antiparticle and a DM fermion respectively; $\alpha, \beta$ are colour indices in the fundamental representation of SU(3) and $p,q=1,2$ spinorial indices. 
The optical theorem relates the matching coefficients, which fix the parameters of this first low-energy theory,  to heavy scalar annihilations into light degrees of freedom, i.e. the in-vacuum cross sections. Neglecting the masses of SM particles  they  read \cite{Biondini:2018pwp}
\begin{eqnarray}
&& c_1=0 \, , \quad c_2= \frac{|y|^2 (|h|^2+ C_F g_s^2)}{128 \pi}\, , \quad c_3 = \frac{1}{32 \pi} \left( \lambda_3 + \frac{C_F g_s^4 }{N_c}\right)  \nonumber \\
&& c_4=\frac{g_s^4}{64 \pi} \frac{N_c^2-4}{N_c} \, , \quad c_5 = \frac{|y|^4}{128 \pi}\, ,
\label{match_coeff}
\end{eqnarray}
where $C_F=(N_c^2-1)/(2 N_c)$ and $h$ stands for the Yukawa coupling between the SM Higgs and the quarks. In this model, the Majorana DM pairs annihilate in two quarks and it features a helicity-suppressed $s$-wave rate \cite{Goldberg:1983nd, Ciafaloni:2011sa, Garny:2015wea}. This is reflected by $c_1=0$ in eq.~\ref{match_coeff} because the masses of the Standard model particles are set to zero. However, in order to treat the boundary between the coannihilation regime and the standard freeze-out correctly an accurate description of $\chi\chi$ annihilations is  highly desirable. Therefore, we will include the leading contribution to $\chi\chi$ annihilations even though it is of higher order in $1/M_\chi$. The situation is different for heavy and light quarks respectively.  In the former case, we compute the matching coefficient for the first four-particle operator in (\ref{Lag_NREFT}) keeping a non-vanishing top-quark mass, and we find $c_1=|y|^4 N_c (m_t/M_\chi)^2 /(128 \pi)$, where $m_t$ is the top-quark mass. In the latter situation, the ratio $m_{u(d)}/M_\chi$ is tiny and we take the $p$-wave contribution as the leading term for $\chi \chi$ annihilations. In the NREFT language, we have to include $1/M_\chi^4$ suppressed operators in (\ref{Lag_NREFT}) which give velocity suppressed contributions to the in-vacuum cross section of order $v^2/M_\chi^2$. When thermally averaging these terms, a $T/M$ suppression appears \cite{Gondolo:1990dk}. The dimension-8 operators and their matching coefficients are given in Appendix~\ref{Appx:higherOperators}. We stress that we do not include the effects of either the top-quark mass or $p$-wave suppressed contributions for $\chi$-$\eta$ coannihilations and coloured scalar pair annihilation as these already possess a non-vanishing contribution at lowest order in $1/M_\chi$.
  
As shown in refs.\cite{Kim:2016zyy,Kim:2016kxt}  the  thermally averaged annihilation cross section can be then written as $\langle \sigma_{{\rm{eff}}} v \rangle = \sum_i c_i \langle \mathcal{O}_i \rangle_T $, where $\langle \cdots  \rangle_T$ stands for the thermal average and $\mathcal{O}_i$ are the four-particle operators listed in eq.~(\ref{Lag_NREFT}). This expression makes the factorization of the scales $M_\chi \gg M_\chi v , M_\chi v^2, \pi T, m_D $ manifest. On a physical ground, hard annihilations happen at a distance scale much smaller than any other introduced by the soft energy scales. These processes are local and unaffected by the medium. 
Soft momenta are taken into account in the thermal average of the very same operators and characterize the dynamics of the coloured scalar before their annihilations: multiple gluon exchange, mass corrections, colour-phase decoherence. A linear response analysis shows that \cite{Kim:2016zyy,Kim:2016kxt} 
\begin{equation}
\langle \mathcal{O}_i \rangle_T = e^{-2M_\chi/T} \left( \frac{M_\chi T}{\pi}\right)^{\frac{3}{2}} \int_{-\Lambda}^{\infty} \frac{d E'}{\pi} e^{-E'/T} \, \rho_i(E') \, ,
\end{equation}
where $E'$ is the  energy of relative motion of the pair after factoring out the center of mass dynamics, and $\Lambda$ is a cutoff restricting the average to the non-relativistic regime, in particular $\alpha_s^2 M \ll \Lambda \siml M $ \cite{Kim:2016kxt}.\footnote{We choose
$\Lambda = 3 \alpha_s M^2$ in our numerical computations and have verified that making it e.g. 2-3 times larger plays no role on our numerical resolution.}
The main advantage of this approach is that the spectral functions $\rho_i$ contain all the dynamical information of the annihilating heavy pair\cite{Kim:2016zyy,Kim:2016kxt}. As far as the coloured scalar are concerned, a repeated gluon exchange leads to the Sommerfeld effect for the scattering states and also to the formation of bound states for the colour-singlet channel. At finite temperature, both possibilities need to be considered simultaneously, and thermally averaged over the whole phase space. After reshuffling some terms we can define thermally averaged  Sommerfeld factors \cite{Kim:2016kxt, Biondini:2017ufr, Biondini:2018pwp} which read as follows:
\begin{equation}
\bar{S}_i=\left( \frac{4 \pi}{M_\chi T} \right)^{\frac{3}{2}} e^{  \frac{2 \Delta M_{T}}{T} }  \int_{-\Lambda}^{\infty} \frac{d E'}{\pi} e^{-E'/T } \frac{\rho_i}{N_i} \, .
\label{sommerfeld_factors}
\end{equation}
where $N_i$ are the number of contractions for each operator.  Note that the $\bar{S}_i$ factors account for all non-perturbative effects in-medium and encode Sommerfeld enhancement/suppression, bound state formation and thermal effects. In terms of the latter quantities, the annihilation cross sections can be written as \footnote{We use the notation adopted in ref.\cite{Biondini:2018pwp}.}
\begin{equation}
\langle\sigma_{{\rm{eff}}} v \rangle  =\frac{2c_1 + 4 c_2 N_c e^{-\Delta M_T/T} + N_c\left[  c_3 \bar{S}_3 + 
  c_4  \bar{S}_4 C_F +2 c_5 \bar{S}_5 (N_c+1) \right] e^{-2\Delta M_T/T}}{\left( 1+ N_c e^{-\Delta M_T/T} \right)^2 }\, ,
\label{cross_mod_2}
\end{equation}
where the thermally averaged Sommerfeld factors for processes with $\chi$ in the initial state are unity since the DM is a singlet under QCD. 

A possible way to determine the spectral functions for the non-relativistic pairs (and then the $\bar{S}_i$ through eq.~(\ref{sommerfeld_factors})) is to solve a plasma modified Schr\"odinger equation. It implements a resummation of repeated interactions between the coloured scalars and the particles from the medium. The spectral function has been shown to correspond to the imaginary part of the Green's function that solves the very same Schr\"odinger equation \cite{Burnier:2007qm,Laine:2007gj}
\begin{eqnarray}
&& \biggl[ 
   H_T -i\Gamma_T(r) - E'
 \biggr] G^{ }_i(E';\bm{r},\bm{r}')  =  
 N^{ }_i\, \delta^{(3)}(\bm{r}-\bm{r}') \, ,
   \label{SH_like_1} 
 \\
 \phantom{s} \nonumber
 \\
&&\lim_{\bm{r},\bm{r}' \to \bm{0}} {\rm{Im}}  G^{ }_i(E';\bm{r},\bm{r}')
  =  \rho^{ }_i(E')  \, ,
  \label{SH_like}
\end{eqnarray} 
where the Hamiltonian has the standard form $H=-\nabla_{\bm{r}}/M_\chi +V_T(r)$ with $r=|\bm{r}|$, $V_T(r)$ is an in-medium potential whereas $\Gamma_T(r)$ represents the interaction rates, i.e.~real scatterings with the plasma constituents (Landau damping and gluon emission/absorption) that has been interpreted as the imaginary part of the heavy-pair potential \cite{Laine:2006ns, Brambilla:2008cx}.  We remark that the potentials are functions of the temperature and their explicit form depends on where the thermal scales sit with respect to the non-relativistic scales (see figure~\ref{fig:figEFT}). The potential and the width in eq.~(\ref{SH_like}) comprise an $r$-independent and $r$-dependent part. The former comes from self-energy diagrams of the each particle of the pair, whereas the latter comes from a gluon attached to the two different heavy particles.
 
The heavy scalar potentials in the different annihilation channels can be interpreted in a rigorous and modern language as matching coefficients of a pNREFT that we discuss in the next section.  
\subsection{QCD-potentials for coloured scalars in pNREFT} 
\label{sec:pNREFT_spectr}
In this section we deal with the non-relativistic heavy-scalar pairs and the QCD potentials affecting them. 
The quantum field theory that describes the energy modes of order $M_\eta v^2$ is a pNREFT. We are interested in such energy modes since we want to study the bound-state formation and dynamics. Well-established examples of effective theory of this kind are pNRQCD for a heavy quark-antiquark pair at zero temperature  \cite{Pineda:1997bj,Brambilla:1999xf} and its generalization to finite temperature \cite{Brambilla:2008cx} (for the Abelian version see refs.\cite{Escobedo:2008sy,Escobedo:2010tu}). In practice one has to integrate out the soft momenta of order $M_\eta v$ characterizing the gluon-momentum transfer between the heavy scalar-antiscalar pair. These modes are indeed still dynamical in the NREFT.\footnote{At this stage it is important to note that the QCD scale $\Lambda_{{\rm{QCD}}}$ has been always assumed to be smaller than any other scales. This way one can perform a perturbative matching from the NREFT onto pNREFT.} Then, depending on whether the temperature scale is larger or smaller than $M_\eta v$, different pNREFTs are obtained (see refs.\cite{Brambilla:2008cx,Brambilla:2010vq, Ghiglieri:2013iya} for heavy quarkonium). In our case, the degrees of freedom are the singlet, octet and sextet wave function fields and the matching coefficients correspond to the singlet, octet and sextet potentials respectively. The presence of the Majorana DM fermion allow for the annihilation channel $\eta \eta \to q q$, so that a particle-particle potential is also relevant.

As mentioned in the previous section, the details of the low-energy theories depend on the assumed hierarchy of scales. 
In an expanding Universe, the temperature of the plasma is decreasing with time and different arrangements of the energy scales are expected to occur. For example, the condition $\pi T \gg M_\eta v$, which is valid around the early stages of the scalar annihilations is not going to hold at later stages and one should plug different potentials in the Schr\"odinger equation (\ref{SH_like_1}). As far as the potential arising from inelastic parton scattering, results are available for quarkonium in the case $M_\eta v \gg \pi T$ \cite{Brambilla:2008cx,Brambilla:2013dpa}, however one would also need the interpolating regime $M_\eta v \approx \pi T$ that is realized during the early Universe cooling. The real part of the potential in the latter situation has not been calculated yet and is subject of a future work \cite{Biondini_preparation}. 
Here, we do not pursue a sophisticated analysis by inserting many different thermal potentials in eq.~(\ref{SH_like_1}) because the focus of this work is on the model phenomenology and the ties with the relic density constraints that capture bound-state formation. In particular, we use the high temperature potentials derived in ref.\cite{Biondini:2018pwp} that rely on the Hard thermal loop (HTL) approximation and account for a Debye-screened Yukawa potential and soft $2 \to 2$ scatterings with particles of the plasma (i.e.~the Landau damping as shown in figure~\ref{fig:figEFT}, diagram $a$). We exploit these potentials for the whole temperature range. Even though this is not strictly rigorous, the corresponding effects are decreasing with the temperature and capture the physics of the thermal environment in the QCD sector. As a refinement with respect to the previous study undertaken in ref.~\cite{Biondini:2018pwp}, we add the gluodissociation contribution for the singlet channel in the thermal width $\Gamma_T(r)$ (process $b$ in figure~\ref{fig:figEFT}) and the accompanying thermal modification in the real part of the potential $V_T(r)$. We use the expressions derived in ref.\cite{Brambilla:2008cx} in the static limit.
In the pNREFT language, this process is induced by chromolectric transitions between singlet and octet fields. It is expected to be subleading at temperatures $\pi T, m_D \gg \Delta V$, where $\Delta V$ is the difference between the singlet and octet energy. Additional details on the pNREFT and the thermal potentials are offered in Appendix~\ref{Appx:thermalpotential}. In this way, we include both the processes relevant for bound state dynamics in a unified framework.

Let us now write down the effective Lagrangian.
In the high-temperature assumption, namely $\pi T \gg Mv$, the temperature scale has to be integrated out first. The main outcome is that the gluon propagator is modified by a thermal mass $m_D$, and also an imaginary part. The latter accounts for real scatterings with the parton constituents of the plasma (quarks and gluons). The gluon propagator takes the so-called Hard Thermal Loop (HTL) form\cite{Pisarski:1988vd,Braaten:1989mz}. Then the typical heavy-scalar momentum transfer can be integrated out and we obtain a pNREFT$_{\rm{HTL}}$ (we use the same language as given in refs.\cite{Vairo:2009ih, Ghiglieri:2012rp}). The degrees of freedom of such EFT are colour singlet, octet and sextet field. The leading contribution to the Lagrangian in the static limit reads 
\begin{eqnarray}
\mathcal{L}_{\hbox{\tiny pNREFT$_{\rm{HTL}}$}}= \mathcal{L}_{{\hbox{\tiny HTL}}} &+& \int d^3 r {\rm{Tr}} \left\lbrace S^{\dagger} \left[ \partial_0 - V_s -\delta M_s\right] S + O^{\dagger} \left[ D_0 - V_o -\delta M_o\right] O \right. 
 \nonumber \\
&&\left. \phantom{xxxxx}+ \,  \Sigma^{\dagger} \left[ D_0 - V_\Sigma -\delta M_\Sigma \right] \Sigma \right\rbrace    \, ,
\label{pNREFT}
\end{eqnarray}
where the trace refers to a colour trace, then $S$, $O$ and $\Sigma$ are the singlet, octet and sextet fields, $D_0=\partial_0 +i g_s A_0^a T^a$, $V_{s(o,\Sigma)}$ are the $r$-dependent potentials whereas $\delta M_{s(o,\Sigma)}$ are the self-energy corrections for the heavy pair and $\mathcal{L}_{{\hbox{\tiny HTL}}}$ is the HTL Lagrangian for the gluons and light quarks. These quantities contain both a real and an imaginary part and read \cite{Laine:2006ns,Brambilla:2008cx, Biondini:2018pwp}   
\begin{eqnarray}
&& V_s(r)= \alpha_s C_F \left[  - \frac{e^{-m_D \, r}}{r} + i  T  \,  \Phi_r(m_D \, r) \right]   \, ,
\label{HTL_singlet}
\\
&& V_o(r)=   \frac{\alpha_s}{2 N_c} \left[  \frac{e^{-m_D \, r}}{r} - i  T \Phi_r(m_D \, r)  \right]  \, ,
\label{HTL_octet}
\\
&& V_\Sigma (r) =   \frac{\alpha_s C_F}{ N_c+1}   \left[  \frac{e^{-m_D \, r}}{r} - i  T \Phi_r(m_D \, r) \right] 
\label{HTL_sextet} \, ,  
\\ \nonumber
\\
&& \delta M_{s}=\delta M_{o}=\delta M_{\Sigma}= -\alpha_s C_F \left( m_D + i T \right) \, ,
\label{HTL_mass}
\end{eqnarray} 
where the auxiliary function $\Phi$ is given by \cite{Laine:2006ns}
\begin{equation}
\Phi(r m_D)=\frac{2}{m_D r} \int_0^\infty \frac{dz}{(1+z^2)^2} \sin (z \, m_D r) \, . 
\end{equation}   
The real and imaginary part of eqs.~(\ref{HTL_singlet}-\ref{HTL_mass}) stand for the $r$-dependent and $r$-independent terms of $V_T(r)$ and $\Gamma_T(r)$ and have to be inserted in the plasma-modified Schr\"odinger equation (\ref{SH_like_1}) for each annihilation channel. 
In contrast to the case of heavy quark-quark pairs studied in \cite{Brambilla:2005yk}, the antitriplet field does not contribute in the case of scalar pairs due to Bose-Einstein statistics.
We would like to stress that even though the potentials have previously been derived in \cite{Biondini:2018pwp}  their interpretation in terms of the pNREFT (\ref{pNREFT}) for heavy coloured scalar comes as a novel aspect in this work. By working with a pNREFT we can make contact with the formalism developed for heavy quarkonium at finite temperature which ($i$) facilitates the identification of the relevant degrees of freedom at the energy scale $M_\eta v^2$ ($ii$) allows a rigorous derivation of the thermal potentials according to the assumed hierarchy of scales ($iii$) provides a  unified framework in which all the processes relevant for bound-state dynamics are included. 

\subsection{Qualitative discussion of relic density}
\begin{figure}[t!]
\centering
\includegraphics[scale=0.74]{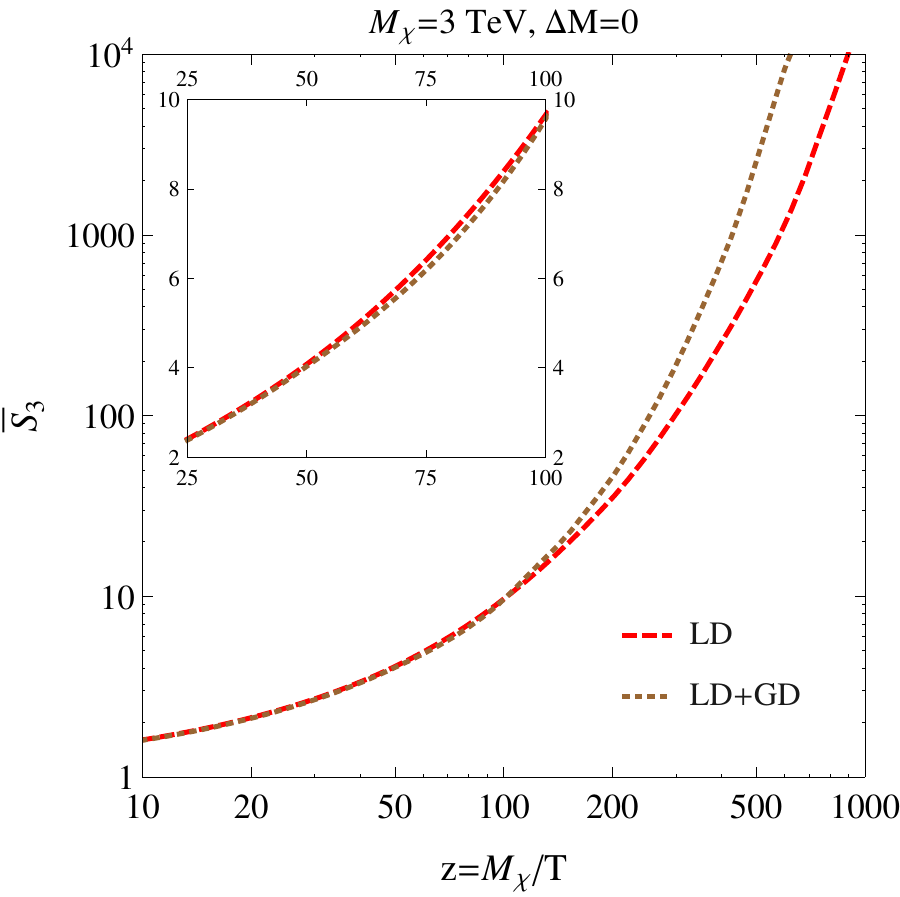}
\hspace{1.0 cm}
\includegraphics[scale=0.69]{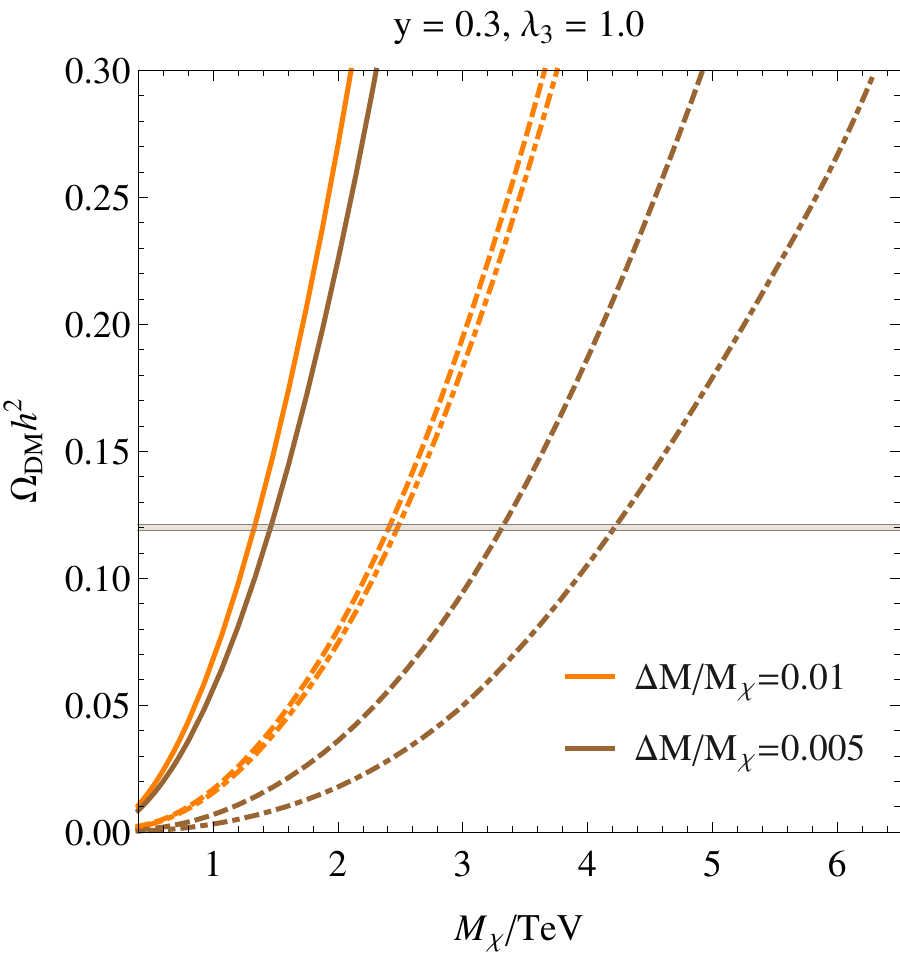}
\caption{\label{fig:fig2}(Left plot) The generalized Sommerfeld factor $\bar{S}_3$  corresponding to the annihilation of the coloured scalars via the singlet channel. The red-dashed lines correspond to the case where only the Landau damping and Debye screening is included, whereas in the brown-dotted line we include the effect of gluo-dissociation/radiative emission.  (Right plot) The DM abundance at
$z_{{\rm{fin}}}=10^3$ where the horizontal gray band shows the observed value $\Omega_{\hbox{\tiny DM}} = 0.1200(12)$\cite{Akrami:2018vks}. The couplings are fixed to $y=0.3$ and $\lambda_3=1.0$ and we show the result for  $\Delta M / M_\chi =0.01$ and $\Delta M / M_\chi =0.005$.  The solid lines shows the free (without gluon exchange) result for the selected $\Delta M / M_\chi$ values, whereas dashed and dot-dashed are obtained with the generalized Sommerfeld factors.}
\end{figure}

Let us briefly discuss our results for the generalized Sommerfeld factors and the corresponding effect on heavy scalar annihilations. In the left panel of figure~\ref{fig:fig2}, we show the generalized Sommerfeld factor $\bar{S}_3$ with and without the effect of gluo-dissociation/bound-state formation by radiative emission and thermal corrections to the real part of the potential for $M_\eta=3$ TeV (see eq.(\ref{GD_potential}) for details). At high temperatures we find a very small effect on the Sommerfeld factor $\bar{S}_3$ that is slightly reduced by the additional dissociation process. On the contrary, at smaller temperatures that corresponds to $z \gsim 300$, the contribution from $O \to S + g $ (octet into gluon and singlet) is sizeable and it helps in forming bound states via the process of the radiative emission as described in other works\cite{Liew:2016hqo,Mitridate:2017izz,Harz:2018csl}.\footnote{In these works the spectral function method is not exploited. Instead, the bound-state formation cross sections calculated by the authors of refs.\cite{Liew:2016hqo,Mitridate:2017izz,Harz:2018csl} are related via the Milne relation to the dissociation (or ionization) cross sections which are obtained from the dissociation rates $\Gamma_T$.}  In this work we focus on $\bar{S}_3$ that enters the thermally averaged cross section in the singlet channel. Here, the bound-state formation boosts the annihilation rate at small temperatures and gives a large effect. In figure \ref{fig:fig2} the Sommerfeld factor $\bar{S}_3$ is shown for $\Delta M = 0$ for illustration. Note, however, that the large enhancement at very small $T$ is deceptive since a non-vanishing mass splitting will cut-off this behavior due to the exponential suppression in the effective cross section~(\ref{cross_mod_2}).\footnote{In the absence of a mass splitting the use of the Boltzmann equation (\ref{BE_gen}) becomes questionable for $T<|E_1'|$, $E_1$ being the binding energy of the lowest-lying bound state, and a more careful determination of the rate equations would be necessary to track the system in this regime~\cite{Binder:2018znk}.} For the repulsive channels, we use the values of $\bar{S}_4$ and $\bar{S}_5$ as derived in ref.\cite{Biondini:2018pwp}. Their effect is to reduce the corresponding octet and sextet annihilation channels, however the effect is less pronounced than in the singlet channel. In our work we do not include the chromoelectric transitions that affect the octet self-energy (there are no transitions of the sextet field because the antitriplet configuration is absent for scalars). Nevertheless this should be pursued for a  more rigorous theoretical treatment and it poses an interesting topic for future research \cite{Biondini_preparation}.\footnote{The octet self-energy in the hierarchy of scales $M_Q v \gg T \gg m_Q v^2 \gg m_D, \, \Lambda_{\hbox{\tiny QCD}}$ has been computed in ref.\cite{Brambilla:2017zei} for heavy quarkonium, where $M_Q$ is the heavy quark mass.}  
In the right panel of figure~\ref{fig:fig2}, the relic density is displayed for $y=0.3$ and $\lambda_3=1.0$ and for selected values of $\Delta M / M_\chi$.  The solid orange and brown lines correspond to the relic density obtained with the free cross section, 
namely without the inclusion of the gluon exchange ($\bar{S}_3=\bar{S}_4=\bar{S}_5=1$ in eq.~(\ref{cross_mod_2})), and for finite mass splitting between the dark matter and the coannihilating specie, $\Delta M /M_\chi=0.01$ and $\Delta M /M_\chi=0.005$. We take the solid lines as reference for the non-perturbation effects that we consider. The dashed lines correspond to a computation  where only the Debye screened Yukawa potential and the thermal width by the inelastic parton scattering is included \cite{Biondini:2018pwp}, whereas the results denoted by  dotted-dashed lines also account for  the $2 \to 1$ process for bound-state dissociation/formation. We notice that the effect is more prominent as the mass splitting decreases. Indeed, the final relic abundance is more sensitive to the coloured scalar annihilations when the scalar particles are very close in mass to the DM fermion.  

Besides the gluon exchange, the Standard Model Higgs boson can induce an attractive potential between heavy coloured scalars in this model. The Higgs exchange has been recently considered in refs.~\cite{Harz:2017dlj,Biondini:2018xor} and the corresponding effect depends on how the interaction between the coloured scalar and Higgs boson is realized.  On the one hand, for a trilinear vertex $h \eta^\dagger \eta$ induced after electroweak symmetry breaking by the Higgs vev (as in the Lagrangian given in \ref{Lag_RT}) the Higgs exchange plays little role \cite{Biondini:2018xor}. The main reason is the heavy mass suppression that, together with the scalar coupling  values considered here $\lambda_3 \leq 1.5$, generates an effective coupling $\alpha_{\hbox{\scriptsize eff}} \equiv \lambda_3 v_h/( 8 \pi M) \approx 0.01$ that is typical of weak interactions strengths for the masses under consideration. Therefore, we neglect Sommerfeld enhancement and bound-state formation induced by the Standard Model Higgs boson in the present work.  On the other hand, a trilinear interaction which is proportional to a new mass scale can have a sizeable impact if the new scale/interaction strength is large enough \cite{Harz:2017dlj}.

\section{Experimental constraints}
\label{sec:exp_constraint}

Naturally, one wonders whether the cosmologically preferred regions of the parameter space can be probed experimentally.  
Collider searches, direct and indirect detection experiments  are known to have a great sensitivity to models with colour-charged mediators  \cite{Bai:2013iqa,An:2013xka,Chang:2013oia,DiFranzo:2013vra,Garny:2014waa}.
In contrast to freeze-out, which is largely insensitive to the flavour of the quark with which the DM interacts, the rates and signatures at colliders and direct detection experiments depend quite sensitively on the specific quark. In order to bracket the range of possibilities we consider two representative choices in the following: DM interactions with (i) the up quark as a representative of the valence quarks and (ii) the top quark.  

\subsection{Collider searches}
\label{sec:LHC}

In collider searches  the impact of the quark flavour on the phenomenology is twofold. First, new production modes which are sensitive to the couplings and masses of the DM can contribute significantly to the total production rate at the LHC if the DM couples to quarks with a large parton luminosity. Second, the flavour of the quark which is produced in the decay of the mediator $\eta$ has a non-negligible impact on the sensitivity of experimental searches.

\subsubsection{Valence quarks}

The production of new physics at colliders receives contributions from different processes. First, gluon-gluon and $q\bar{q}$ initial states allow for the production of  $\eta \eta^\dagger$ pairs by gluon exchange due to the colour charge of the mediator. The production amplitude from quarks-antiquark pairs also includes a term with t-channel $\chi$ exchange which can dominate over the QCD production for large values of $y$. In addition, $\eta \eta$ ($\eta^\dagger \eta^\dagger$) and $\eta \chi$ ($\eta^\dagger \chi $) final states are produced exclusively by new physics. Same sign scalar pair production, which is possible because of the Majorana nature of $\chi$, is mediated by  t-channel $\chi$ exchange. In the regime where new physics contributions to scalar pair production are relevant the $\eta \eta$ rate is typically enhanced compared to $\eta \eta^\dagger$ and $\eta^\dagger \eta^\dagger$ due to the large $u$-quark parton distribution function (PDF) and dominates over the other channels for large $y$ \cite{Garny:2014waa}. In addition, (anti-)quark gluon initial states allow for the production of $\eta^{(\dagger)} \chi$ final states. The amplitude of this production mode is linear in the new physics coupling $y$ and therefore most important at intermediate values of $y$ since the t-channel processes scale with a higher power of $y$.   

At colliders the final states of models with colour charged mediators consist of jets from the decay of the mediator and pairs of DM particles. Since the DM can not be observed with the LHC detectors $\chi$ production leads to a momentum imbalance of the observed particles and the signal consists of missing transverse energy ($MET$)  in association with jets.  In the coannihilation regime the mass different $\Delta M $ is small and the jets from $\eta$ decay are typically soft which makes them hard to distinguish from QCD backgrounds. Consequently, hard partons in the final state are necessary to provide additional handles on the event and  suppress the backgrounds. In this case the DM will recoil against the visible jets (or other SM particles) and thus increase the  missing transverse energy. Numerous searches for these types of signatures exist, \cite{Aaboud:2017phn,Sirunyan:2017jix,Sirunyan:2017ewk,Sirunyan:2017hnk,Aaboud:2016obm,Aaboud:2017dor} with monojet searches typically being the most sensitive.

We implement the simplified model in FeynRules \cite{Alloul:2013bka} and simulate the collider signal with MadGraph5$\_$aMC@NLO \cite{Alwall:2014hca}. The results are then passed to Pythia8 \cite{Sjostrand:2014zea} for showering and hadronization. Detector effects are taken into account with Delphes 3 \cite{deFavereau:2013fsa} and we use the Checkmate 2 \cite{Dercks:2016npn} implementation of the ATLAS monojet search \cite{Aaboud:2017phn} \footnote{Technically the Checkmate implementation is based on a preliminary version of the ATLAS monojet search \cite{ATLAS-CONF-2017-060}.  The presentation differs from the published version but the underlying experimental results are the same.} to derive limits on the cosmologically favoured parameter space.

Our analysis shows that the pure QCD production cross section of $\eta$ mass-degenerate with $\chi$ is excluded for $M_\eta < 320$~GeV. In the mass degenerate regime the new physics production modes increase the reach of the LHC search to $540$~GeV for $y=1$ and $1100$~GeV for $y=2$. However, the relic density constraint imposes a non-negligible mass splitting between the DM and $\eta$ for these couplings. As a consequence, the region of parameter space which is tested by the ATLAS search does not overlap with the region of parameter space considered in our analysis of the relic density. 

\subsubsection{Top Quarks}

For $\eta$ interacting with top quarks the story is rather different. First, the parton luminosity of  top quarks is completely negligible at the relevant center of mass energies. Consequently, only the pure QCD production of the mediator is relevant in this case.  The decay of the mediator is more complicated than in the case of light quarks. In the region of parameter space for which coannihilations are relevant $m_t + M_\chi > M_\eta$ such that the two-body decay $\eta \rightarrow \chi t$ is kinematically not accessible. Therefore, three-body and four-body decays into a b-quark, an on- or off-shell W boson and $\chi$ dominate the width  and the finals states become rather complex compared to $\eta$ with couplings to light quarks. 
The production and decay modes of $\eta$ are similar to the simplified topologies used in searches for supersymmetry  \cite{Sirunyan:2017cwe,Sirunyan:2017wif,Aaboud:2017aeu,Aaboud:2017phn} and the results for stops  apply to the model considered here. Currently, the LHC is only sensitive to $M_\chi \lesssim 500$~GeV and the regions of parameter space of interest remains unconstrained. 

\subsection{Direct detection}

\label{sec:DirectDetection}
Direct detection experiments probe the scattering rate of DM particle off nuclei in low-background detectors. 
In the last years the sensitivity of direct searches has seen a huge improvement and the best current limits probe scattering cross sections as low as $4. 1 \times 10^{-47}\, \mbox{cm}^2$ \cite{Aprile:2018dbl}. As a consequence these experiments are now able to test scenarios which feature a suppressed scattering rate such as coannihilations. 

In the model under consideration here, the direct detection cross section is rather sensitive to the fundamental interactions between the DM and the quarks.  On the one hand, DM interacting with light quarks couples  directly to the valence quarks in the nucleus and a sizeable interaction is already generated at tree-level. On the other hand, the tree-level interactions between DM coupling to heavy quarks and the nucleons are highly suppressed since the expectation value of heavy quark pairs in protons and neutrons is negligible. Consequently loop-induced couplings between the DM and light quarks and gluons become relevant and dominate the DM nucleon cross section in this case.
In general both spin-dependent (SD) and spin-independent (SI) interactions are generated. However, due to the higher experimental sensitivity SI DM-nucleon scattering dominates the constraints and we have checked that  SD scattering is not competitive. Therefore, we will focus on SI interactions in the following. 

In general, the SI direct detection cross section on nucleons is given by~\cite{Jungman:1995df}
\begin{align}
\sigma_{SI}^N =\frac{4}{\pi}\frac{M_\chi^2 m_N^2}{(M_\chi+m_N)^2} f_N^2\;,
\end{align}
where $N$ denotes the nucleon, i.e.~proton or neutron,  $m_N$ is the nucleon mass and $f_N$ is the effective DM nucleon coupling. The value of $f_N$ depends on the underlying interactions between the DM and the constituents of the nucleons, i.e. valence quarks and gluons.
The relevant terms in the effective Lagrangian and their contribution to $f_N$ are rather different if DM couples to light or heavy quarks. To make the discussion more transparent we will therefore discuss the two cases separately.

Before discussing direct detection  in detail it should be noted, however, that the limits and prospects of direct searches depend on details of the astrophysics of DM which are not fully understood. In our work we assume the standard halo model, i.e.~a truncated Maxwellian velocity distribution, with $v_0=220 \mbox{km}/\mbox{s}$, $v_{esc}=544 \mbox{km}/\mbox{s}$ and $v_{e}=232 \,\mbox{km/s}$ and taken the local DM density to be $\rho_0= 0.3 \,\mbox{GeV}/\mbox{cm}^3$ in agreement with the standard choice made by the experimental collaborations for the presentation of direct detection limits \cite{Aprile:2018dbl}.

\subsubsection{Valence quarks}

\begin{figure}[t]
\centering
\includegraphics[width=0.5\textwidth]{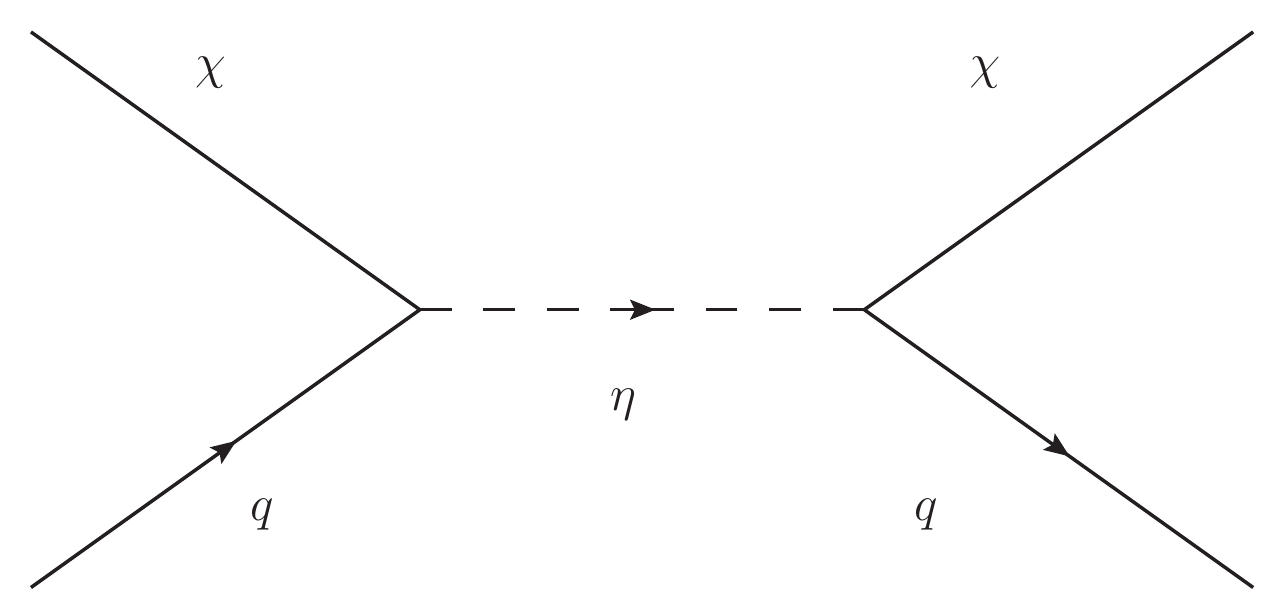}
\caption{Representative diagram contributing to the effective DM nucleon interaction via tree-level $\eta$ exchange. \label{fig:tree}}
\end{figure}

If the DM interacts directly with light quarks the dominant contribution to the effective DM nucleon coupling arises from the tree-level exchange of $\eta$, see Fig.~\ref{fig:tree} for a representative diagram. In addition, the loop-induced coupling between the DM and the Higgs can become relevant for large $\lambda_3$  and this lead to an additional contribution to the DM nucleon coupling. Taking all contributions into account one finds~\cite{Drees:1993bu,Hisano:2010ct,Garny:2015wea} 

\begin{align}
\left. \frac{f_N}{m_N}\right.\biggl|_{\hbox{\scriptsize valence}} =&-\sum_{q=u,d,s} f_{Tq}^N \left( \frac{M_\chi g_q}{2} - \frac{g_{h\chi \chi}}{2 v_h m_h^2}\right)+ f_{TG}^N \frac{2}{9} \frac{g_{h \chi \chi}}{2 v_h \, m_h^2} - \sum_{q=u,d,s} (3 q(2)+3 \bar{q}(2)) \frac{M_\chi g_q}{2} 
\label{eq:effDMnucleon_light}
\end{align}
where $m_h$ is the SM Higgs mass. The $f^N_{Ti}$ are related to nucleonic matrix elements of the quark
and gluon operators by $f^N_{Tq}= \langle N|m_q \bar{q} q| N \rangle/m_N$ and $f_{TG}=1-\sum_{q=u,d,s} f_{Tq}^N$  and parametrize the contributions
of quarks and gluons to the nucleon mass.  The quantities $q(2)$ and $\bar{q}(2)$ denote the second moments of the quark and anti-quark parton distribution functions, respectively. In our numerical analysis  we use the default values  for $f_{Tq}$ from micrOMEGAs 5.0 \cite{micromegasmanual} while $q(2)$ and $\bar{q}(2)$ are taken from \cite{Hisano:2010ct}. The fundamental interaction of the DM with partons is encapsulated in the parameters $g_q$ and $g_{h\chi \chi}$.

The first term in Eq.~\ref{eq:effDMnucleon_light} arises from an effective four fermion interaction $\chi \chi \bar{q}q$ between the DM and the quarks. For DM coupling to light quarks this interaction is generated by tree-level $\eta$ exchange with a strength given by \cite{Drees:1993bu}
\begin{align}
g_q=-\frac{y^2}{8}\frac{1}{(M_\eta^2-(M_\chi+m_q)^2)^2}
\,.
\end{align}
In addition, triangle diagram with $\eta$ and the quarks running in the loop induced an effective DM Higgs coupling $g_{h\chi\chi}$. Higgs exchange contributes to the $\chi \chi \bar{q} q $  interaction and   induces a coupling to the gluons once heavy quarks are integrated out.  Due to the small Higgs coupling to valence quarks only the diagrams with the Higgs coupling to $\eta$ contribute appreciable to $g_{h\chi\chi}$ and the interaction with the quarks can be neglected.
In the zero-momentum transfer limit relevant for direct detection the effective coupling is given by
\begin{align}
g_{h \chi \chi}=\frac{3\lambda_3 y^2}{16 \pi^2} \frac{v_h}{M_\chi} \left((x_\eta-1) \log \left(\frac{x_\eta}{x_\eta-1}\right)-1 \right) \;,
\end{align}
where $x_\eta=M_\eta^2/M_\chi^2$ and the quark masses have been neglected.
 This contribution is always subdominant for DM  but can amount to a $\mathcal{O}(10\%)$ correction  in certain regions of parameter space if $\lambda_3$ is large.

The last term in eq.~\ref{eq:effDMnucleon_light} is also generated by tree-level $\eta$ exchange and the strength of the interactions is controlled by the same parameter $g_q$ which enters in the first term.

\subsubsection{Top quark}

\begin{figure}
\centering
\includegraphics[width=0.35\textwidth]{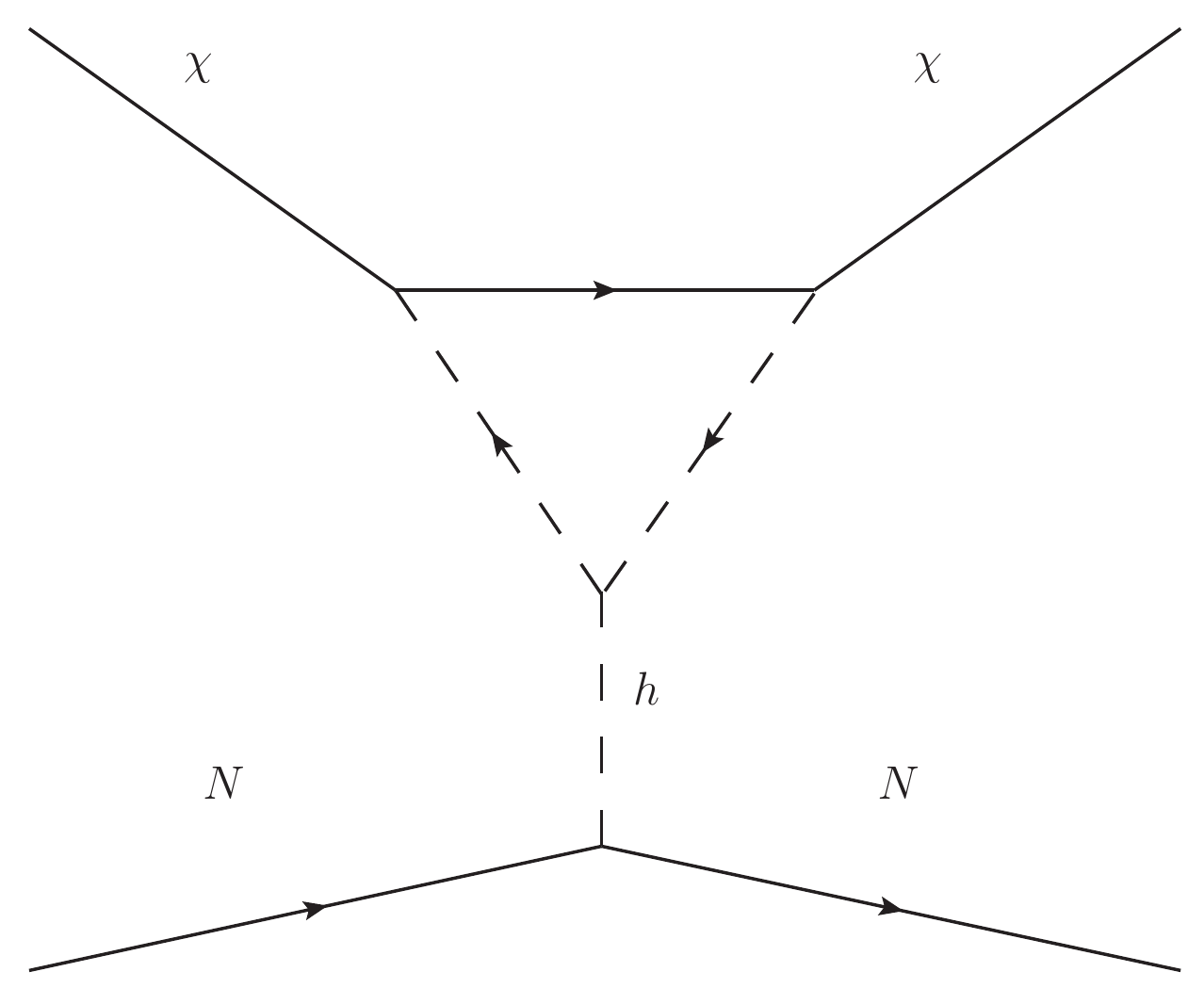}
\hspace{1.2 cm}
\includegraphics[width=0.35\textwidth]{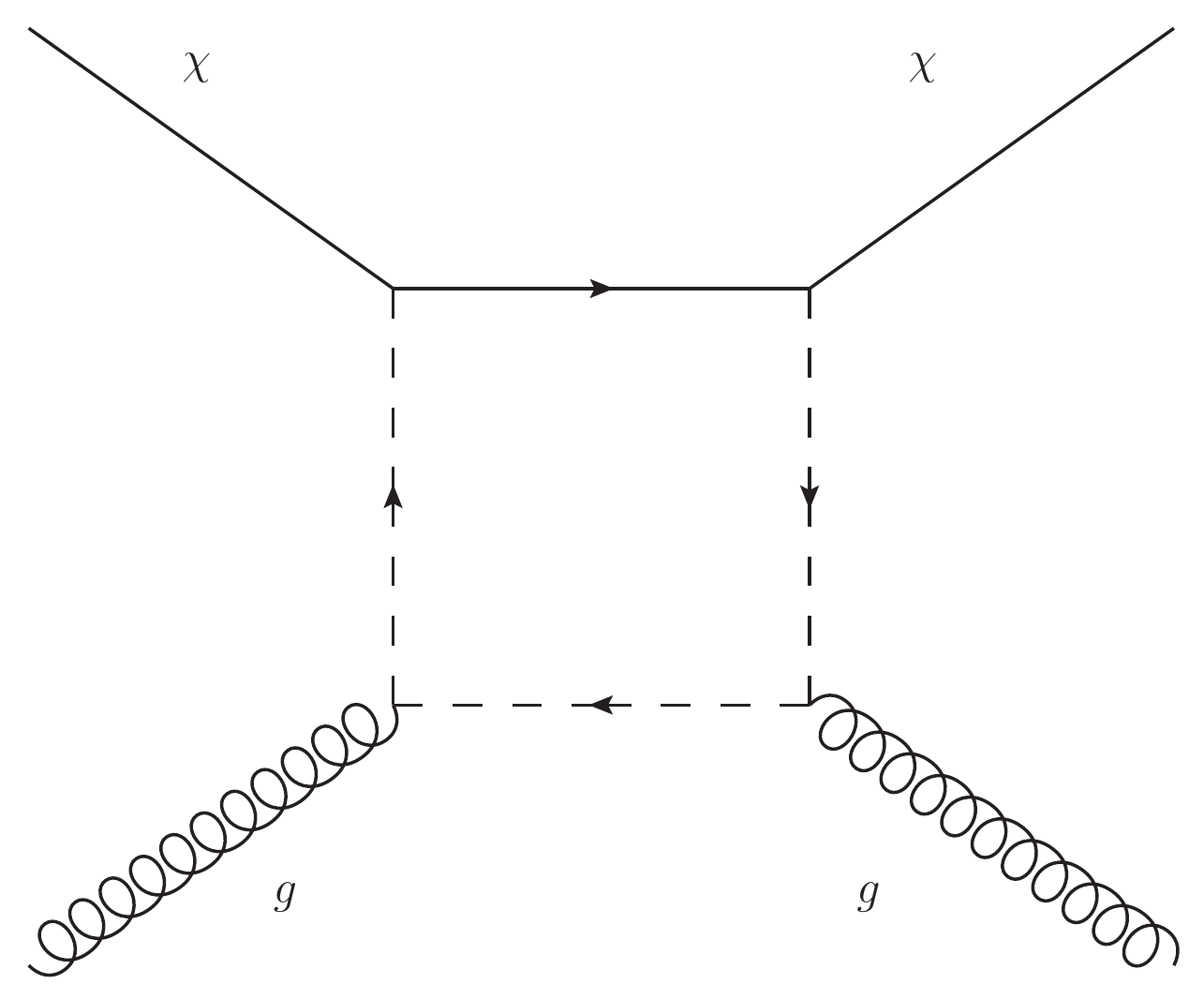}
\caption{Exemplary one-loop diagrams contributing to the effective DM nucleon coupling. The triangle diagram (left) contributes to the direct detection rate through an effective coupling to the Higgs while the box (right) leads to an interaction with the gluons content of the nucleons.\label{fig:loop}}
\end{figure}

For DM interacting with heavy quarks, the direct tree-level coupling to the nucleus is absent since the abundance of heavy quarks in the nucleons is highly suppressed. Consequently, higher order processes become important. Two terms contribute substantially to the effective DM nucleon interaction: (i) the effective DM Higgs coupling leads to an interaction via Higgs exchange; (ii) box loops with quarks and $\eta$ in the loop generate an effective coupling between the DM and the gluon content of the nucleons, see Fig.~\ref{fig:loop} for representative diagrams contributing to the loop induced coupling.

In the low energy limit, the ensuing DM nucleons coupling reads~\cite{Drees:1993bu,Hisano:2010ct,Garny:2015wea} 
\begin{align}
\left. \frac{f_N}{m_N} \right.\biggl|_{\hbox{\scriptsize top}}=&\sum_{q=u,d,s} f_{Tq}^N  \frac{g_{h\chi \chi}}{2 v_h m_h^2} +f_{TG}^N \frac{2}{9} \frac{g_{h \chi \chi}}{2 v_h m_h^2}  
-f_{TG}^N \frac{8\pi}{9\alpha_S} b + \frac{3}{4} G(2) g_G\, ,
\label{eq:effDMnucleon_heavy}
\end{align}
where $G(2)$ is the second moment of the gluon parton distribution while $b$ and $g_G$ parametrize the strength of addition contributions to the DM nucleon interaction which were negligible for valence quarks. 

The first two terms arise from the effective DM Higgs coupling and are similar to the ones introduced in the previous section. Due to the large Yukawa coupling of heavy quarks, additional diagrams with the Higgs coupling to the internal quarks line contribute to $g_{h\chi\chi}$. Taking the full quark mass dependence into account the effective coupling in the zero-momentum transfer limit reads~\cite{Ibarra:2015nca} 
\begin{align}
g_{h\chi \chi}&=\frac{3y^2}{8 \pi^2}\frac{m_t^2}{v_h M_\chi}\mathcal{A}^t+\frac{3 y^2 \lambda_3}{32\pi^2}\frac{v_h}{M_\chi} \mathcal{A}^\eta\;, 
\end{align}
where 
\begin{align}
\mathcal{A}^t&= \frac{1}{2}(x_t-x_\eta)\log\left(\frac{x_\eta}{x_t}\right)+\frac{x_\eta(x_\eta-2x_t-1)+x_t(x_t-1)}{D}\log\left(\frac{x_\eta +x_t+D-1}{2\sqrt{x_\eta}\sqrt{x_t}}\right)+1  \, ,\\
\mathcal{A}^\eta&=(x_\eta-x_t-1)\log\left(\frac{x_\eta}{x_t}\right) -\frac{2((x_\eta-x_t-1)^2-2x_t)}{D}\log\left(\frac{x_\eta+x_t+D-1}{2\sqrt{x_\eta}\sqrt{x_t}}\right) -2  \, ,
\end{align}     
with $D=\sqrt{(x_\eta-x_t-1)^2-4x_t}$ and $x_t=m_t^2/M_\chi^2$.

The remaining terms arise from box-diagrams which generate an effective coupling between the DM and the gluon content of the nucleus. The third term is induced by the $\chi \chi G^a_{\mu \nu} G^{a \mu \nu}$ operator where $G^a_{\mu \nu}$ denotes the field strength tensor of QCD. The strength of this interaction is given by \cite{Drees:1993bu}
\begin{align}
b=\frac{\alpha_S y^2}{8 \pi} M_\chi \left(\frac{1}{8} I_2 -\frac{M_\chi^2}{12} I_4 -\frac{1}{24} I_5 \right)\;.
\end{align}
The $I_i(M_\eta,m_q,M_\chi)$ are comparatively lengthy loop functions which are given in the Appendix of\cite{Gondolo:2013wwa}.

 The box diagram also contributes to an operator which connects the DM to the gluon twist-2 operator, i.e. the traceless part of the energy momentum tensor
 \begin{align}
  \mathcal{O}_{twist}=\chi (i \partial^\mu) (i \partial^\nu)\chi \left(  G_\mu^{a \rho} G^a_{\rho \nu} + \frac{1}{4} g_{\mu \nu} G^{a}_{\alpha \beta} G^{a \alpha \beta} \right)       \,.
\end{align}  The coefficient of the effective Lagrangian is given by \cite{Drees:1993bu} 
\begin{align}
g_G= \frac{\alpha_s y^2}{24 \pi} M_\chi \left(M_\chi^2 I_4 +\frac{1}{2} I_5 \right)\,.
\end{align}  

\subsection{Indirect detection}

Indirect detection experiments measure the flux of cosmic rays produced by dark matter annihilations.
Today  coannihilation processes are no longer active and only the annihilations of dark matter pairs still proceed. Therefore, the annihilation rate of coannihilating DM in the Universe today is suppressed  with respect to the cross section expected for a generic thermal relic.  The present day annihilation cross section $\sigma v$  is given by the $c_1$ piece in Eq.~\ref{cross_mod_2}. 
As noted previously, $c_1$ vanishes at lowest order and the leading contribution is either suppressed by $m_q^2/M_\chi^2$ or by $v^2$ depending on the quark mass. For dark matter coupling to the top quark this is only a mild suppression provided that $M_\chi$ is not too large.
Analyzing the parameter space preferred by the observed relic abundance we find that the largest present day annihilation cross section is $\sigma v (\chi \chi \rightarrow t \bar{t}) = 7.4 \times 10^{-27} \; \mbox{cm}^3/\mbox{s}$.  Current indirect searches using Fermi-LAT data \cite{Fermi-LAT:2016uux} or antiprotons \cite{Cuoco:2017iax,Reinert:2017aga} are not sensitive to such a small cross section for $M_\chi \geq 500$~GeV and do not constrain the parameter space. This situation may improve in the future but a recent analysis has shown that probing $\sigma v \lesssim 2\times 10^{-26} \; \mbox{cm}^3/\mbox{s}$ will even be a challenge for CTA \cite{Balazs:2017hxh}.
 For dark matter coupling to light quarks  the situation is more severe and, taking $v_{fo}$ as the typical velocity at freeze-out, we expect a $v^2/v_{fo}^2\approx 10^{-5}$ suppression of the annihilation rate today compared to the rate at freeze-out.\footnote{The  velocity of dark matter in astrophysical structures is bound by the escape velocity. In our Galaxy the typical velocity is   $v/c \approx 10^{-3}$.}
This suppressions is so strong that three-body final states $q \bar{q} B$, where $B$ is a SM boson, dominate the annihilation rate today since they are s-wave and helicity unsuppressed  \cite{Bergstrom:1989jr,Garny:2011cj,Ciafaloni:2011sa,Bell:2010ei,Luo:2013bua}. However, their present day annihilation rate is still reduced compared to two-body rate at freeze-out due to the higher power in the SM coupling constant and the three-body phase space. Consequently, the cosmologically preferred annihilation rate is also out of reach of continuum gamma-ray and charged cosmic ray searches in this case. Interestingly, the spectrum of the bosons in the $q \bar{q} B$ final states is strongly peaked. For $B=\gamma$ this leads to a line-like feature which can be used to devise a more sensitive gamma-ray search \cite{Bergstrom:1989jr,Bringmann:2012vr}. Unfortunately, no clear evidence for a gamma ray lines has been detected so far and bounds based on Fermi-LAT and H.E.S.S. data do not test thermal dark matter in models similar to the one considered here \cite{Garny:2013ama,Clark:2019nby}.

\section{Results}
\label{Sec:Results}

In this section we combine our results for the abundance of thermal relics with the limits from direct detection experiments and LHC searches.
Since the relic abundance is known very precisely we can fix one of the model parameters ($y$) in terms of the others ($M_\chi, M_\eta, \lambda_3$) by imposing $\Omega_{\hbox{\tiny DM}} h^2 = 0.1200(12)$\cite{Akrami:2018vks}. 
In order to make the discussion more transparent we consider different slices of the parameter space and  discuss their phenomenology. To be more specific, we choose representative values for $\lambda_3$, i.e. $\lambda_3=0$ and $\lambda_3=1.5$, which bracket the impact of $\lambda_3$ and allows for an intuitive visualization of the results. In our analysis we focus on $\Delta M / M_\chi \leq 0.2$ since coannihilations are subdominant at larger mass splitting and we restrict ourself to $M_\chi \geq 500$~GeV. For lower $M_\chi$ the late stage annihilations due to bounds-state formation make the freeze-out process sensitive to temperatures at which the QCD potentials become non-perturbative. In this regime our results are no longer reliable and one would need to determine the thermal expectation values on a lattice. Such an analysis goes beyond the scope of this work.   
As in the previous section, we first discuss the phenomenology of light quarks before turning to top quarks. In order to avoid repeating similar results we limit ourself to DM coupling to up quarks in the following and do not discuss the other light quarks in detail.

\begin{figure}
\centering
\includegraphics[width=0.49\textwidth]{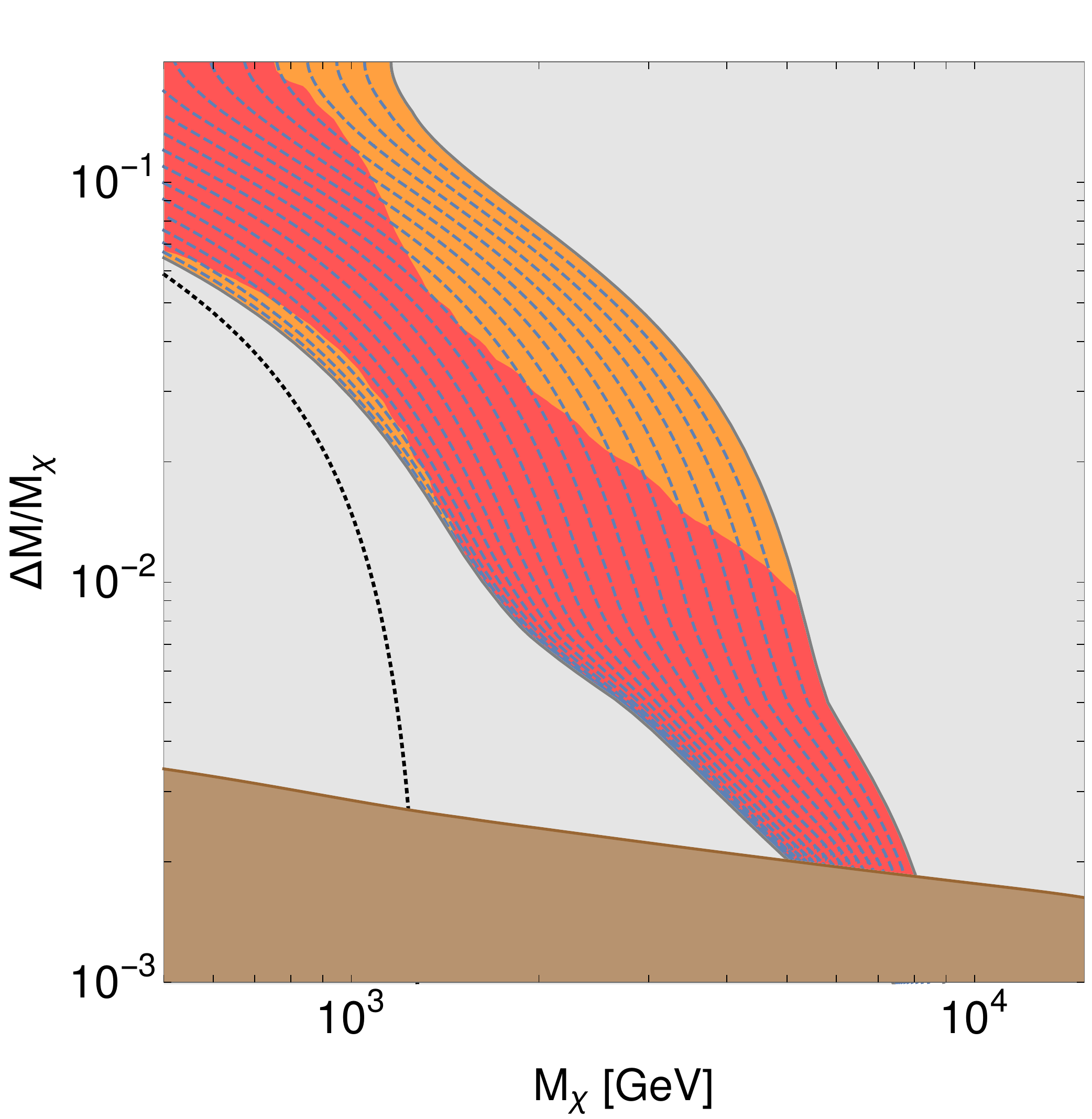}
\includegraphics[width=0.49\textwidth]{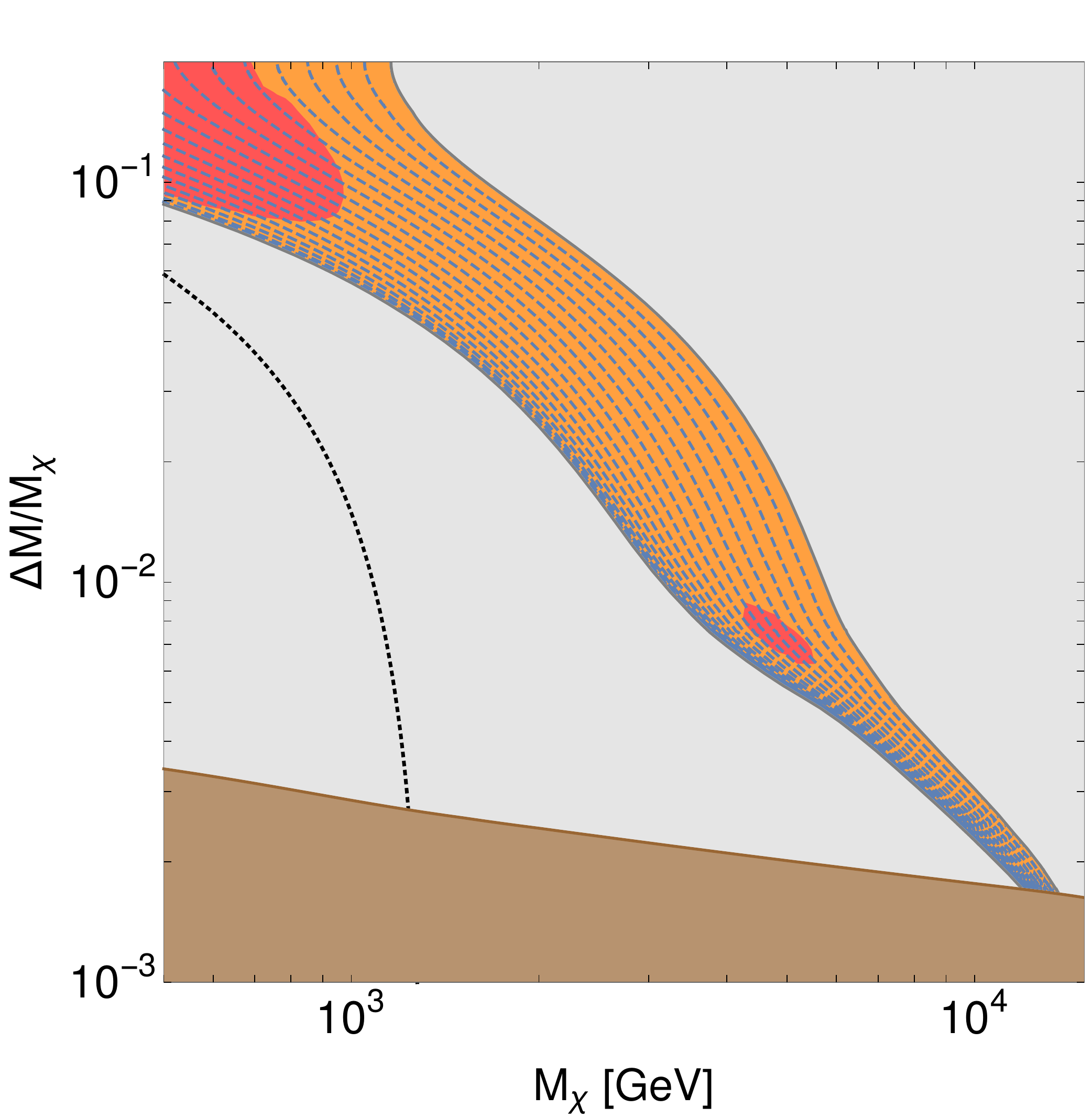}
\caption{The left (right) panel summarized constraints on a thermally produced DM coupling to $u$-quarks via a coloured mediator $\eta$ for $\lambda_3=0.0$ ($\lambda_3=1.5$). The red region is excluded by results of the Xenon1T experiment. The orange region shows the part of parameter space which can be tested by a future Darwin-like direct detection experiment. In the upper gray area $y\geq2$ is required to reproduce the correct relic abundance. In the lower gray region the thermal abundance is controlled exclusively the gauge interactions and the observed relic abundance can not be achieved for DM from freeze-out. In the brown region, twice the mass splitting is smaller than the binding energy of the lightest $\eta \eta^\dagger$ bound state. The dashed blue lines indicate the required value of $y$ while the dotted black line indicates results obained with the in-vacuum cross section and $y=0.3$.\label{fig:summary_up}}
\end{figure}
 
\begin{figure}
\centering
\includegraphics[width=0.45\textwidth]{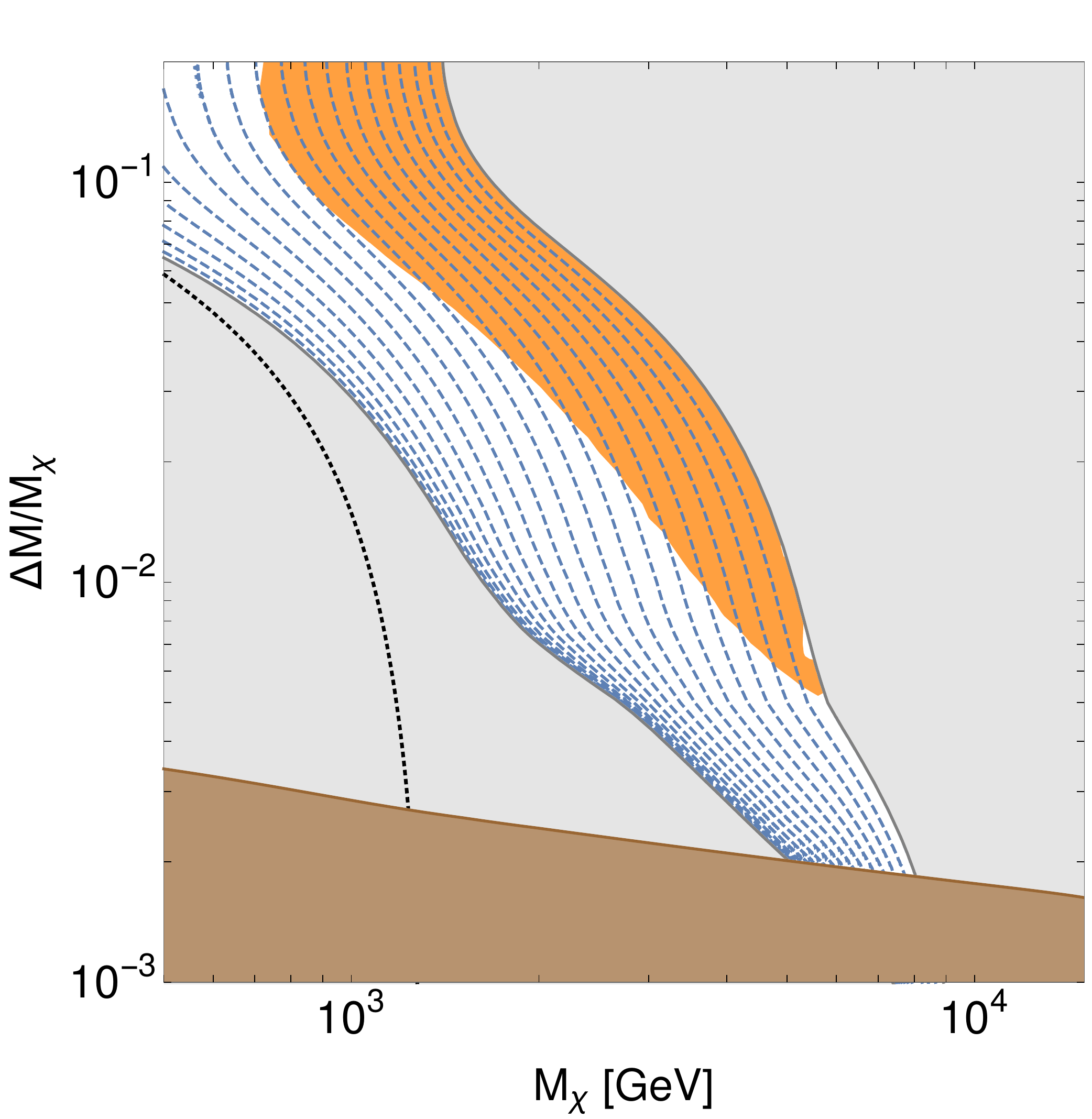}
\includegraphics[width=0.45\textwidth]{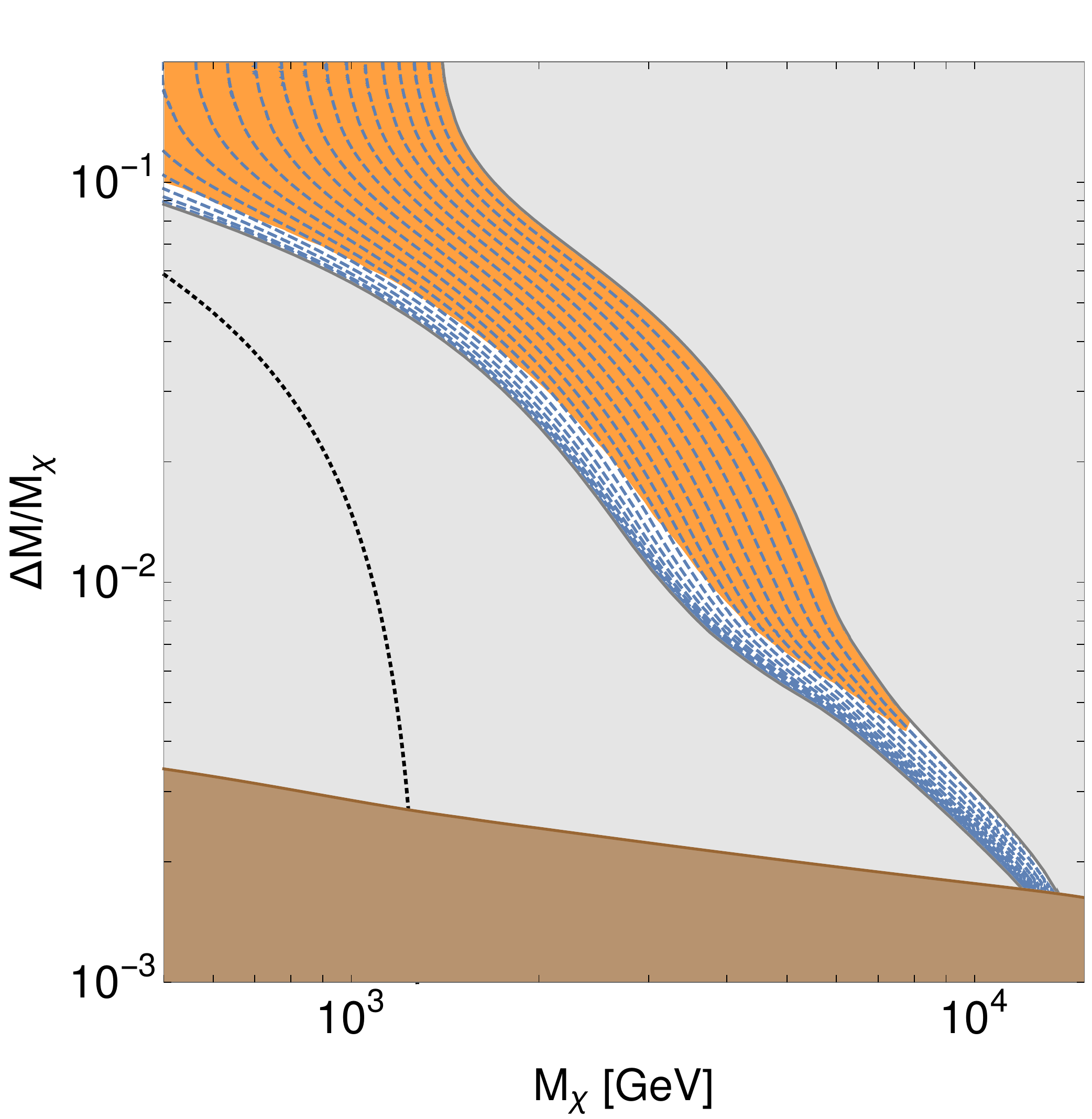}
\caption{Same as Fig.~\ref{fig:summary_up} for DM coupling to top quarks. Note that the coupling $y$ preferred by thermal DM is the same as in Fig.~\ref{fig:summary_up} since the quark mass (Higgs Yukawa coupling)  does not have an appreciable effect on the relic density in the part of parameter space shown here. \label{fig:summary_top}}
\end{figure}

As can be seen in Fig.~\ref{fig:summary_up} the parameter space for successful freeze-out is limited from above and below.  The blue dashed curves reproduce the observed relic density in the $(M_\chi, \Delta M / M_\chi)$ plane for representative values of $y$ ranging from $y=0.1$ to $y=2.0$ in steps of $0.1$. They are obtained via the numerical solution of the Boltzmann equation for the DM yield given in eq.~(\ref{Boltzmann_Y}). The effective annihilation cross section $ \sigma_{{\rm{eff}}}  v$ is defined in eq.~(\ref{cross_mod_2}) and it includes the dynamics of coloured mediators in medium, i.e. Sommerfeld enhancement,  bound-state formation/dissociation and thermal effects as explained in Sec.~\ref{sec:NREFT_spectr} and \ref{sec:pNREFT_spectr}. On the one side, as $\Delta M /M_\chi $ increases coannihilations become less efficient and the value of $y$ which reproduced $\Omega_{\hbox{\tiny DM}} h^2=0.1200(12) $ increases until the limit is reached. On the other side, for low  $\Delta M /M_\chi $ coannihilations become very efficient and since some of the coannihilation rates are exclusively set by the Standard Model charges of the mediator we find a region (lower gray area) in which the coannihilation rate is too large too accommodate $\Omega_{\hbox{\tiny DM}} h^2=0.1200(12)$ irrespectively of $y$.\footnote{For very small values of $y$ the assumptions entering the coannihilation formalism, namely chemical equilibrium between $\eta$ and $\chi$, break down and production of DM through the conversion-driven freeze-out mechanism becomes possible \cite{Garny:2017rxs}.} 
As can be seen the slope of the relic density curves changes noticeably at high and at low $\Delta M / M_\chi$. The first change at $\Delta M / M_\chi$ in the range $0.1 \-- 0.2$ corresponds to the onset of non-negligible coannihilation and marks to upper edge of the parameter space in which the in-medium effects considered here are relevant.
 The change of slope at $\Delta M/M \approx 5 \times 10^{-3}$ in Fig.~\ref{fig:summary_up} and Fig.~\ref{fig:summary_top} is due to very efficient bound-state formation that corresponds to large coloured scalar annihilation rate at small temperatures. This was anticipated in Fig.~\ref{fig:fig2} where we show the relic abundance for different values of  $\Delta M/M$. The smaller the mass splitting the longer the coloured scalars are as abundant as the DM and, therefore, the abundance of the overall dark sector is driven by large annihilation rates of the coloured particles.  For comparison, the back dotted line indicates the exemplary results obtained with a pure in-vacuum computation that neglects Sommerfeld and bound state effects for $y=0.3$.
By increasing $\lambda_3$ to a value close to the perturbative limit (we show results for $\lambda_3=1.5$) the coannihilation processes get markedly more efficient and the region of freeze-out DM shifts upwards in the $M_\chi,\Delta M/M_\chi $ plane. 
This is due to the fact that the Higgs contributes to the singlet channel annihilation rate which is enhanced by  the especially large $\bar{S}_3$ Sommerfeld factor in the thermally averaged cross section (\ref{cross_mod_2}).
In the brown region, the binding energy of the lowest-lying bound states is larger than two times the in-vacuum mass splitting, i.e.~$2 \Delta M < |E_1'|$, such that the lightest two-particle states in the dark sector
are the bound states formed by the coloured scalars. Due to efficient chemical equilibration rates that convert $\eta$ particles into DM particles, it seems plausible that almost all  DM fermions convert into the scalars and
subsequently annihilate away, so that the model is most likely not able to explain the viable 
observed DM abundance in this parameter region \cite{Biondini:2018pwp}.  

Collider searches are not  able to probe a part of the cosmologically favoured part of the parameter space for coannihilating DM. The ATLAS monojet search is not sensitive enough the exclude the pure QCD-monojet cross section, i.e. $\eta$ pair production in association with a jet, for  $m_\eta \geq 320$ GeV. The additional production modes due to new physics which depend on $y$ enhance the monojet cross section substantially and dominate the monojet rate for large $y$ but even in this case the cosmologically preferred $\eta$ masses for a given $y$ are outside of the reach of the LHC search by several hundred GeV.    
 
Direct detection experiments can test the coannihilation scenario and, for $\lambda_3=0$, a big part the parameter space is already excluded by the null-results reported by XENON1T~\cite{Aprile:2018dbl}. In addition, the prospects for testing this scenarios with future experiments are excellent and a detector with an exposure similar to the  proposed DARWIN experiment\cite{Aalbers:2016jon} could probe the entire region. However, the larger mass splitting associated with a larger value of $\lambda_3$ decrease the DM nucleon cross section and makes these scenarios harder to access experimentally. Nevertheless, even for the case of $\lambda_3=1.5$ current experiments are already starting to exclude parts of the high and low mass region and not even the tip of the coannihilation strip, i.e. $M_\chi\geq 10$ TeV, remains beyond the reach of future experiments.   

In the case of DM interacting with top quarks the picture is somewhat different. On the one hand, the cosmological preferred parameter space is essentially insensitive to the quark flavour and remains unchanged.   On the other hand, the collider and the direct detection phenomenology is quite different. As detailed in Sec.~\ref{sec:LHC} collider searches are currently not sensitive to $M_\chi \geq 500$~GeV irrespectively of the other model parameters and, therefore, the LHC limits do not appears in Fig.~\ref{fig:summary_top}. Also direct detection experiments currently lack the sensitivity necessary to probe the parameter space of thermal DM coupling to tops. However, this situation will improve substantially once future experiments begin to collect data. A DARWIN-like detector can test a significant part of the parameter space. In contrast to the case DM coupling to up quarks the direct detection prospects for DM coupling to tops become better as $\lambda_3$ increases and  for $\lambda_3=1.5$ it is almost possible to reach $M_\chi= 10$ TeV  for large values of $y$. This is due to the fact that the loop mediated DM Higgs coupling, which is sensitive to $\lambda_3$, contributes significantly to the direct detection rate while the suppression due to the higher mass splitting associated with large $\lambda_3$ is less pronounced than for valence quarks. 
All considered the prospects for thermal DM produced by coannihilations with a colour charged mediator in the future are encouraging and strengthen the case for an ultimate direct detection experiment such as DARWIN.

\section{Conclusions and Outlook}
\label{Sec:Conclusions}

In this paper we connect the recent improvements for the relic density determination of WIMPs  with current  experimental limits and future prospects. To be specific, we consider a simplified model with a  Majorana fermion as DM candidate and a strongly interacting scalar. The latter acts as a mediator between the Standard Model and the dark sector and is assumed to be close in mass with the DM in order to allow for significant coannihilations. 
In this case, the DM relic density is influenced by the annihilations of the coloured scalars that feel strong interactions. Due to repeated gluon exchanges and interactions with light plasma constituents in the thermal medium, the determination of the coloured-scalar annihilation rates is non-trivial and requires care. In the last years a great effort has been made to include the effects of Sommerfeld enhancement and bound-state formation in the relic density computations. These new results indicate that previous analyses underestimated the impact of coannihilation on the relic density and 
 call for a reassessment of the experimental capabilities  and a detailed study of the parameter space preferred by thermal freeze-out.  

It is manifestly subtle and complicated to include bound-state dynamics in a thermal medium due to an intricate interplay between non-relativistic and thermal energy scales. 
In this work, we adopt a formalism that connects the thermally averaged cross section with the  determination of the spectral function for non-relativistic particle pairs. The latter can be extracted by solving a plasma-modified Schr\"odinger equation. In doing so, one can describe the two-particle state entering the hard annihilation and account for the dynamical formation of bound states in a thermal bath, without any assumption on the nature of the annihilating pair. As far as the coloured scalar pairs are concerned, different processes for bound-state dissociation/formation are active: $2 \to 2$ soft scattering with plasma constitutes and gluo-dissociation/absorption. These are captured by the imaginary part of the thermal potentials experienced by the coloured scalars. In addition, the real part of the potential is modified by the thermal plasma. Profiting from recent progress in the context of heavy quarkonia in a hot QCD medium, we use a pNREFT for heavy coloured scalars in the high-temperature limit. Here, the thermal potentials for the singlet, octet and sextet fields are understood as matching coefficients, and account for Debye screening and inelastic parton scattering in the real and imaginary part respectively. 

In order to attain a more complete description of the coloured scalar dynamics and make contact with other results in the literature, we have also taken into account the process of bound-state dynamics triggered by a thermal gluon. In the pNREFT language, it is described by singlet to octet transitions. When extracting the generalized Sommerfeld factors we find results which agree with the general arguments provided by the EFTs for heavy quarks in a medium and the corresponding power counting. Singlet-to-octet transitions give a substantial contribution for small temperatures (here identified with $z \gsi 300$) whereas they plays little role at higher temperatures. To the best of our knowledge, both the processes responsible for bound-state dynamics have not been considered simultaneously in this model  before.  

The thermal potentials among non-relativistic particles are a key ingredient to a better understanding of DM dynamics in a thermal bath. However, it is also important to focus on the rate equation governing the time evolution of the DM abundance in order to connect it with phenomenology. Here, we used a Boltzmann equation and plug in it a thermally average cross section that tracks the nature of the two particles annihilating: scattering or bound states depending on the temperature and the other model parameters.

With these ingredients a precise determination of the DM relic abundance is possible and the cosmologically favoured regions of the parameter space can be mapped out.  
We focus on regions with strong coannihilations, i.e. $\Delta M/ M \leq 0.2$, and restrict ourself to $M_\chi \geq 500$~GeV. An analysis of $M_\chi < 500$~GeV would be of considerable interest, however, for such low DM masses QCD can enter the non-perturbative regime before the freeze-out process is complete. The formalism that we use can handle the non-perturbative regime of QCD, however one would have to evaluate the thermal expectation values of the heavy particles on a lattice.
The preferred regions of parameter space can then be confronted with direct detection and collider experiments. Unfortunately, we find that current LHC searches are not sensitive to the region of parameter space of interest here due to the small $y$/high $M_\eta$ required to reproduce the observed relic density.  In contrast to collider searches, direct detection experiments have an excellent sensitivity to coannihilating DM. In models of DM interacting with light quarks current bounds on the DM nucleon scattering cross section already exclude significant parts of the parameter space and future detectors such as the proposed DARWIN experiments will be able to probe the complete parameter space conclusively. The situation is a bit less optimistic if DM interacts preferably with heavy quarks and current experiments are not yet able to exclude DM interactions with top quarks. Nevertheless, even in this case future experiments will be able to probe a significant fraction of the parameter space.

Finally, we would like to stress that  there remains room for improvements in the determination of the thermal relic abundance. First of all, a more detailed classification of the thermal potentials according to different assumption on the scale hierarchies would be highly desirable and the impact of changing scale arrangements during the evolution of the system ought to be explored on a quantitative level. Moreover, the dynamics induced by chromoelectric transitions for the colour octet configuration have not been included so far and an assessment of the corresponding effect on the repulsive thermal potential could improve the precision of the relic density calculation further. Recently, an alternative derivation of the rate equations for the DM number density has been carried out~\cite{Binder:2018znk}. In this case, an ab initio out-of-equilibrium treatment is pursued and a useful systematization of the different rates governing the DM dynamics can be handled. The general equations can be simplified according to different assumptions and, for example, the result obtained from a linear response theory is recovered as a limiting regime close to chemical equilibrium. Perhaps, another interesting and promising approach to a detailed dynamics of color-singlet and -octet densities is the one discussed in refs. \cite{Brambilla:2016wgg,Brambilla:2017zei,Blaizot:2017ypk}. Here the heavy quark-antiquark pair is interpreted as an open quantum system in a thermal environment and an out-of-equilibrium dynamics is also taken as a starting point. It would be worth exploring this direction for applications to relic density determination for coloured coannihilators and compare quantitatively the outcomes of the different theoretical approaches.

\section*{Acknowledgements}
This work of S.B. was supported by the Swiss National Science Foundation (SNF) under grant
200020-168988. S.B. thanks Nora Brambilla, Mikko Laine, Kalliopi Petraki  and Antonio Vairo for helpful discussions. 

\appendix
\numberwithin{equation}{section}
\section{$1/M^4_\chi$ operators}
\label{Appx:higherOperators}
In this section we discuss the four-particle operators responsible for velocity-suppressed $\chi \chi$ annihilations in the NREFT. Using the language adopted in NRQCD\cite{Bodwin:1994jh} the dimension-8 terms in the Lagrangian are 
\begin{eqnarray}
\mathcal{L}_{\hbox{\tiny NREFT}}^{d=8}= i \left\lbrace  \frac{f(^3 P_0)}{M_\chi^4} \mathcal{O}(^3 P_0) + \frac{f(^3 P_1)}{M_\chi^4} \mathcal{O}(^3 P_1) + \frac{f(^3 P_2)}{M_\chi^4} \mathcal{O}(^3 P_2) + \frac{f(^1 S_0)}{M_\chi^4} \mathcal{O}(^1 S_0) \right\rbrace ,
\end{eqnarray}
where the included operators read  
\begin{eqnarray}
&& \mathcal{O}(^3 P_0) = \frac{1}{3} \psi^\dagger \left( -\frac{i}{2}  \Ddouble \cdot \bm{\sigma} \right) \psi^\dagger \, \psi  \left( -\frac{i}{2}  \Ddouble \cdot \bm{\sigma} \right) \psi \, ,
\label{op_p1}
\\
&& \mathcal{O}(^3 P_1) = \frac{1}{2} \psi^\dagger \left( -\frac{i}{2}  \Ddouble \times \bm{\sigma} \right) \psi^\dagger \, \psi  \left( -\frac{i}{2}  \Ddouble \times \bm{\sigma} \right) \psi \, ,
\label{op_p2}
\\
&& \mathcal{O}(^3 P_2) = \psi^\dagger \left( -\frac{i}{2}  \DdoubleBIS \phantom{}^{(i} \sigma^{j)} \right) \psi^\dagger \, \psi   \left( -\frac{i}{2}  \DdoubleBIS \phantom{}^{(i} \sigma^{j)} \right) \psi \, ,
\\
\label{op_p3}
&& \mathcal{O}(^1 S_0) = \frac{1}{2} \left[\psi^\dagger  \psi^\dagger  \psi \left( -i \frac{\Ddouble}{2} \right)^2 \psi  + {\rm{h.c.}} \right] \, .
\label{op_p4}
\end{eqnarray}
The arguments of the $^{2S+1}L_{J}$ indicate the angular momentum state of the $\chi \chi$ pair which is created or annihilated by the operator, $\Ddouble$ is the difference between the derivative acting on the spinor to the right and on the spinor to the left, and $T^{(ij)}$  stands for the symmetric traceless component of a tensor, i.e.~$T^{(ij)}=(T^{ij}+T^{ji})/2-T^{kk} \delta^{ij}/3$, that appears in eq.~(\ref{op_p3}) through the derivative and Pauli matrices vector components. At variance with NRQCD there is no covariant derivative because the DM Majorana fermion is a singlet under QCD. Moreover, the number of independent operators is smaller than NRQCD because of the relation $\sigma^k_{pq} \sigma^k_{rs} = 2\delta_{ps} \delta_{qr}-\delta_{pq} \delta_{rs}$ as noticed in ref.\cite{Biondini:2018pwp}. Another difference is that a Majorana fermion pair is even under charge conjugation and following the discussion in ref.\cite{Weiler:2013hh} one can already select which operators are allowed or not by looking at the corresponding $J^{PC}$.  

\begin{figure}[t!]
\centering
\includegraphics[scale=0.53]{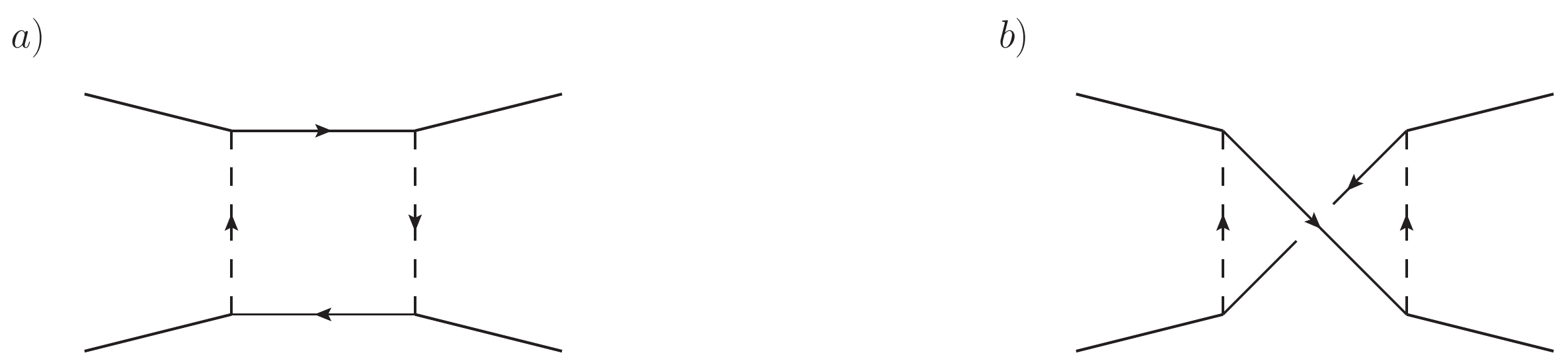}
\caption{\label{fig:Majorana_match_EFT}One-loop diagrams for the matching of the dimension-6 and dimension-8 operators in the NREFT.}
\end{figure}

The matching is done by computing the one-loop diagrams in the fundamental theory given in figure~\ref{fig:Majorana_match_EFT}. One has to expand and keep terms up to order $(v/M_\chi)^2$ and match onto the corresponding amplitude in the NREFT. As mentioned in the body of the paper, we only need the imaginary part of the one-loop amplitude. We perform the calculation in the center of mass frame for the incoming $\chi \chi$ pair. The incoming Majorana fermions have momenta $\bm{p}$ and $-\bm{p}$, whereas the outgoing states have $\bm{p}'$ and $-\bm{p}'$. By energy conservation we have $|\bm{p}|=|\bm{p}'|$ and the velocities are defined as $\bm{v}=\bm{p}/E$ and $\bm{v}'=\bm{p}'/E$, with $E=\sqrt{p^2+M_\chi^2}$.   The result reads (we drop the spinors and we use the notation as in ref.\cite{Bodwin:1994jh})
\begin{eqnarray}
{\rm{Im}}(\mathcal{M}_a + \mathcal{M}_b)=\frac{|y|^4 N_c}{128 \pi M_\chi^2} \left[ \frac{8}{45} \bm{v} \cdot \bm{v}' \, \sigma^i \otimes \sigma^i + \frac{1}{15} \bm{v} \cdot \bm{\sigma} \otimes \bm{v}' \cdot \bm{\sigma} + \frac{1}{15} \bm{v'} \cdot \bm{\sigma} \otimes \bm{v} \cdot \bm{\sigma} \right] \, .
\end{eqnarray} 
This quantity has to be matched with
\begin{eqnarray}
{\rm{Im}} \mathcal{M}_{\hbox{\tiny NREFT}}&=&  \frac{1}{M_\chi^2} \left[ \frac{f(^3 P_1)+f(^3 P_2)}{2} \, \bm{v} \cdot \bm{v}' \, \sigma^i \otimes \sigma^i + \frac{f(^3 P_0)-f(^3 P_2)}{3} \,  \bm{v'} \cdot \bm{\sigma} \otimes \bm{v} \cdot \bm{\sigma} \right.  \nonumber \\
 &&\phantom{xxx}+  \left. \frac{f(^3 P_2)-f(^3 P_1)}{2} \,  \bm{v} \cdot \bm{\sigma} \otimes \bm{v}' \cdot \bm{\sigma}  \right] \, ,
\end{eqnarray}
therefore we obtain
\begin{equation}
f(^3 P_0) = \frac{|y|^4 N_c}{288 \pi} \, , \quad  f(^3 P_1) = \frac{|y|^4 N_c}{1152 \pi} \, , \quad  f(^3 P_2) = \frac{11 \, |y|^4 N_c}{5760 \pi} \, , \quad f(^1 S_0)=0.
\end{equation}
In this calculation the quark mass is set to zero. The number of contractions of the dimension-8 operators contain two powers in the velocity.  When computing the corresponding thermally averaged cross section, one obtains a thermal integral that appears in the velocity expansion (i.e. non-relativistic expansion) of the cross section \cite{Gondolo:1990dk}. Finally it reads \cite{Garny:2015wea, Biondini:2018pwp} 
\begin{eqnarray}
\langle\sigma_{{\rm{eff}}} v \rangle  &=& \frac{3}{2} \frac{T}{M} \left( \frac{2}{3} f(^3 P_0) + 2 f(^3 P_1)+ \frac{10}{3} f(^3 P_2) \right)  + \mbox{coannihilation terms}.
\label{cross_mod_3}
\end{eqnarray} 
This is the thermally averaged cross section that we use to obtain the relic density constraints in the case of DM interacting with light quarks.
\section{EFTs and thermal potentials}
\label{Appx:thermalpotential}
Here we give more details about the numerical implementation of the potentials to extract the spectral functions and the corresponding generalized Sommerfeld factors. As mentioned in the body of the paper, we do not pursue a detailed analysis of the various arrangement for the non-relativistic and thermal scales. We restrict to a simplified scheme with the aim to include the two different processes that describe bound-state formation/dissociation and the corresponding effects on DM annihilations.

We highlight the similarities and differences with respect to the numerical derivation and setting adopted in ref.\cite{Biondini:2018pwp}. First, for the numerical evaluation of the Schr\"odinger equation one needs  $\alpha_s$ for a wide range of energies and the Debye mass for a large range of temperatures. We remind the reader to the appendix of ref.\cite{Biondini:2018pwp} for details on how these quantities are fixed since we stick to the same implementation. We just notice that even going down to temperatures smaller than 1 GeV, we can rely on a perturbative determination of the potentials. Our observation is based on a recent lattice study \cite{Bazavov:2018wmo} that suggests the absence of string-tension contributions for temperatures $T > 2 \, T_c$, where $T_c \approx 150$ MeV  is the QCD-crossover temperature.  Since we consider the smallest DM matter mass to be $M_\chi=0.5$ TeV and we integrate the Boltzmann equation down to $z=10^3$, we are within such a case.

Second, as mentioned in the body of the paper, we add the contribution of the gluo-dissociation process in $V(r)_T$ and $\Gamma(r)_{T}$ for a better estimation of the thermally averaged Sommerfeld factors. In the pNREFT language, this amounts at considering singlet to octet transitions mediated by a chromoelectric gluon. We take the following expression as calculated in ref.~\cite{Brambilla:2008cx} that are valid for the scalar field $\eta$ in the fundamental and antifundamental representation of SU(3)$_c$ 
\begin{equation}
\delta V_{\hbox{\tiny GD}} = \frac{4}{3} C_F \frac{\alpha_s}{\pi} r^2 \, T^2 \Delta V f (\Delta V / T) \, , \quad  \Gamma_{\hbox{\tiny GD}} =  \frac{2}{3} C_F \alpha_s r^2 (\Delta V)^3 n_B (\Delta V),
\label{GD_potential}
\end{equation}
where 
\begin{equation}
f(x)=\int_0^{\infty} dt \frac{t^3}{e^t-1} P \frac{1}{t^2-x^2} \, , \quad n_B(x)=\frac{1}{e^{x/T}-1} \, .
\end{equation}
Here, P stands for the principal value, $n_B$ is the Bose-Einstein distribution and $\Delta V$ is the difference between the singlet and octet static potentials $\Delta V \approx N_c \alpha_s /(2 r)$. For a numerical estimate of $\Delta V$, we used the inverse Bohr radius for the inverse distance, $1/r=1/a_0=M \alpha_s C_F /2$ and then  $\Delta V \approx M  \alpha_s^2$.    
In eq.~(\ref{GD_potential}) the typical expansion of pNRQCD appears, namely the small distance $r$ \cite{Pineda:1997bj,Brambilla:1999xf,Brambilla:2008cx}. When solving the Schroedinger-like equation (\ref{SH_like}) with the addition of the terms in eq.~(\ref{GD_potential}), we impose them to contribute when $1/r \geq \pi T$ is satisfied. This is needed to justify the usage of that potential in first place and also to avoid spurious effects in the spectral function determination at large distances.
Finally, we notice that the Coulomb potential at small distances (and temperatures) is already recovered in the HTL expressions by summing the $r$-dependent with the $r$-independent part in eqs.~(\ref{HTL_singlet}) and (\ref{HTL_mass}). Despite it is not rigorous to use the HTL potentials at small temperatures, namely $\pi T < 1/r$, they still provide a well-defined behaviour of the solution at both large and small distances.

\bibliographystyle{hieeetr}
\bibliography{Majorana_DM.bbl}

\end{document}